\documentclass[a4paper,11pt]{article}

\usepackage{import}
\usepackage{preamble}

\include{preamble.sty}

\title{\boldmath Measurement of the weak mixing angle at a Super Charm-Tau factory with data-driven monitoring of the average electron beam polarization}

\author[]{A. Bondar}
\author[]{A. Grabovsky}
\author[]{A. Reznichenko}
\author[]{A. Rudenko}
\author[]{V. Vorobyev}

\affiliation[]{Budker Institute of Nuclear Physics,\\
11, Acad. Lavrentieva pr., Novosibirsk, 630090 Russia}

\affiliation[]{Novosibirsk State University,\\
2, Pirogova str., Novosibirsk, 630090 Russia}

\emailAdd{A.E.Bondar@inp.nsk.su}
\emailAdd{A.V.Grabovsky@inp.nsk.su}
\emailAdd{A.V.Reznichenko@inp.nsk.su}
\emailAdd{A.S.Rudenko@inp.nsk.su}
\emailAdd{V.S.Vorobev@inp.nsk.su}

\abstract{A method for measuring the average longitudinal polarization of the electron beam at an electron-positron collider operating near the~$\jpsi$ resonance is proposed. The method utilizes the differential cross-section of~$\jpsi\to [\Lambda\to p\pi^-][\lambar\to\pbar\pi^+]$ decay. It can be used to measure the average longitudinal polarization of electrons with the statistical precision better than~$10^{-3}$ at a Super Charm-Tau factory operating at the luminosity of~$10^{35}~\lumi$. The method is discussed in the context of the weak mixing angle measurement in the same experiment.}

\begin{document} 
\maketitle
\flushbottom

\section{Introduction}\label{sec:intro}
Precision experiments at low-energy electron-positron colliders CESR(-c), BEPC, BEPC~II, PEP-II, and KEKB, referred to as flavor factories, yielded a rich harvest of fundamental results in nearly all parts of particle physics: hadron spectroscopy, $CP$ symmetry breaking, physics of $\tau$-lepton, dynamics of strong decays etc. The research program of flavor factories is highly complementary to the physics program of the energy frontier experiments at LHC and must be continued with state-of-the-art particle collider and detector technologies. The new generation of experiments --- super flavor factories --- are going to be on stage in the upcoming years. Super $B$-factory SuperKEKB has already started data acquisition. Two projects of Super Charm-Tau (SCT) factories are under consideration (see refs.~\cite{sct, stc}).

Both SCT factory projects consider the longitudinal polarization of the electron beam at the collision point. Presence of polarized electrons enriches the physics program and provides access to new observables. In particular, studies of tau lepton and baryons gain benefit. 

The central part of the SCT experimental program with the polarized beam is precision electroweak physics. Parity-violating interaction of the $Z$ boson with leptons leads to dependency of $e^+e^-\to\jpsi$ cross section on the helicity of the electron due to the interference of $e^+e^-\to\gamma^\ast\to c\barr{c}$ and~$e^+e^-\to Z^\ast\to c\barr{c}$ processes. The left-right asymmetry, 
\begin{equation}\label{eq:ahel}
 \ahelo \equiv \frac{\sigrt - \siglt}{\sigrt + \siglt},
\end{equation}
is sensitive to this effect. Here~$\sigrt$ and~$\siglt$ are the total~$\jpsi$ production cross sections with right-handed and left-handed electrons, respectively. To the leading order, Standard Model predicts the value (see ref.~\cite{khriplovich})
\begin{equation}\label{eq:jpsi_ahelo}
 \ahelo = \frac{-\sinthw+3/8}{2\sinthw(1-\sinthw)}\left(\frac{m_\jpsi}{m_Z}\right)^2
 \approx 4.7\cdot 10^{-4},
\end{equation}
where~$m_\jpsi=3096.9~\mev$ is the~$\jpsi$ meson mass, $m_{Z}=91.19$~GeV is the~$Z$ boson mass, and~$\thw$ is the effective weak mixing angle that depends on the momentum transfer.  The value of~$\sinthw$ was measured with relative precision of~$0.1\%$ on the $Z$ resonance at LEP and SLC (see ref.~\cite{lepslc}). In a number of experiments~$\sinthw$ was also measured at lower energy transfer with relative precision of a few percent (see refs.~\cite{kumar, erler} for a review). The value~$\sinthw\approx 0.23$ was used to get the estimate in~eq.~(\ref{eq:jpsi_ahelo}).

Electrons are never fully polarized in an experiment. Hence the asymmetry~$\ahelo$ is scaled down to the visible asymmetry~\ahel by the average longitudinal polarization of electrons~$\pe$ ($-1 \leq \pe \leq 1$):
\begin{equation}\label{eq:jpsi_ahel}
 \ahel \equiv \frac{\sigma_{\scriptscriptstyle \pe} - \sigma_{\scriptscriptstyle -\pe}}{\sigma_{\scriptscriptstyle \pe} + \sigma_{\scriptscriptstyle -\pe}} = \ahelo\pe.
\end{equation}

The value of~$\sinthw$ at the~$\jpsi$ energy scale differs from the value at the~$Z$ peak by about~$3\%$. It means that $\sinthw$ should be measured %at least
with sub-percent precision to witness the shift reliably.

Two experimental inputs are necessary to measure~$\sinthw$: the cross section asymmetry~$\ahel$ and the average polarization~$\pe$. Error propagation in eqs.~\eqref{eq:jpsi_ahelo} and~\eqref{eq:jpsi_ahel} leads to the following relation for the relative uncertainties
\begin{equation}\label{eq:uncert}
 \frac{d(\sinthw)}{\sinthw} = C_{\scriptscriptstyle \ahel}\frac{d\ahel}{\ahel} \oplus C_{\scriptscriptstyle \pe}\frac{d\pe}{\pe},
\end{equation}
where the operator~$\oplus$ denotes the square root of sum of squares and
\begin{equation}\label{eq:uncert_coef}
 C_{\scriptscriptstyle \pe} = -C_{\scriptscriptstyle \ahel} = 
 \frac{\left(1-\sinthw\right)\left(3 - 8\sinthw\right)}
 {3\left(1-\sinthw\right) - \sinthw\left(3 - 8\sinthw\right)}
 \approx 0.44.
\end{equation}
Eqs.~\eqref{eq:uncert} and~\eqref{eq:uncert_coef} imply that the relative uncertainties of about~$1\%$ are required for both quantities~$\ahel$ and~$\pe$ to measure $\sinthw$ precisely enough to detect deviation of~$\thw$ from the value at the~$Z$ peak. 

Let us address the expected statistical precision of the~$\ahel$ measurement. Assuming beam energy spread of~$10^{-3}$, the visible~$\jpsi$ production cross section  is (see ref.~\cite{tanya})
\begin{equation}\label{eq:jpsicsec}
 \sigma(e^+e^-\to \jpsi) \approx 3\times 10^{-30}~\mathrm{cm}^2.
\end{equation}
Targeted SCT luminosity of~$10^{35}~\lumi$ will provide about~$10^{12}$~$\jpsi$ mesons detected during a~$10^7~\mathrm{s}$ long period of data taking. Presume the data set is divided into three equal parts containing~$N_0\approx 3\cdot 10^{11}$ events each, corresponding to 1) beam with $+\pe$ average polarization, 2) beam with $-\pe$ average polarization, and 3) unpolarized beam. Assuming~$\pe=0.8$, the statistical uncertainty for the asymmetry~$\ahel$ reads
\begin{equation}\label{eq:ahel-stat}
 \frac{d\ahel}{\ahel} \approx \left[\ahel\sqrt{2N_0\varepsilon}\right]^{-1}
 \approx 5\cdot 10^{-3},
\end{equation}
where the fraction~$\varepsilon$ of~$\jpsi$ decays used in the analysis is estimated to be~$0.5$. The estimate in~eq.~\eqref{eq:ahel-stat} gives  about~$3\cdot 10^{-3}$ for the relative uncertainty of~$\sinthw$. Such a level of precision is comparable with that of the LEP and SLD results. Control of the systematic uncertainties of the measured~$\ahel$ and~$\pe$ is expected to be the major challenge of this experiment. 

The electron beam polarization should be monitored with a dedicated device in real-time during the data taking. However, it is difficult to obtain precise value of the average polarization~$\pe$ from the real-time measurements despite the large statistics. An alternative approach is to measure the average electron beam polarization~$\pe$ directly via analysis of the data collected by detector. This approach is optimal from the systematic uncertainty control viewpoint since the data used for polarization measurement is exactly the same as is used for the~$\ahel$ measurement. The average polarization measurement technique based on analysis of~$\Lambda$-hyperon decays is developed in this work.

The rest of the text is structured as follows: the~$5D$ differential cross section of the~$e^+e^-\to[\Lambda\to p\pi^-][\lambar\to \pbar\pi^+]$ process is derived in section~\ref{sec:matrix_element}; proof-of-concept study is presented in section~\ref{sec:feas};~$\sinthw$ measurement strategy at SCT factories is discussed in section~\ref{sec:systematics}; conclusion is given in section~\ref{sec:conc}. Details of the matrix element calculation are summarized in appendix~\ref{app:matrix}. The angular distributions in the center-of-mass frame are obtained in appendix~\ref{app:cmsdistrib}.

%%%%%%%%%%%%%%%%%%%%%%%%%%%%%%%%%%%%%%
%%%%  Differential cross section  %%%%
%%%%%%%%%%%%%%%%%%%%%%%%%%%%%%%%%%%%%%
\section{Differential cross section} \label{sec:matrix_element}
\begin{figure}[tb]%
 \centering \includegraphics[width=.8\textwidth]{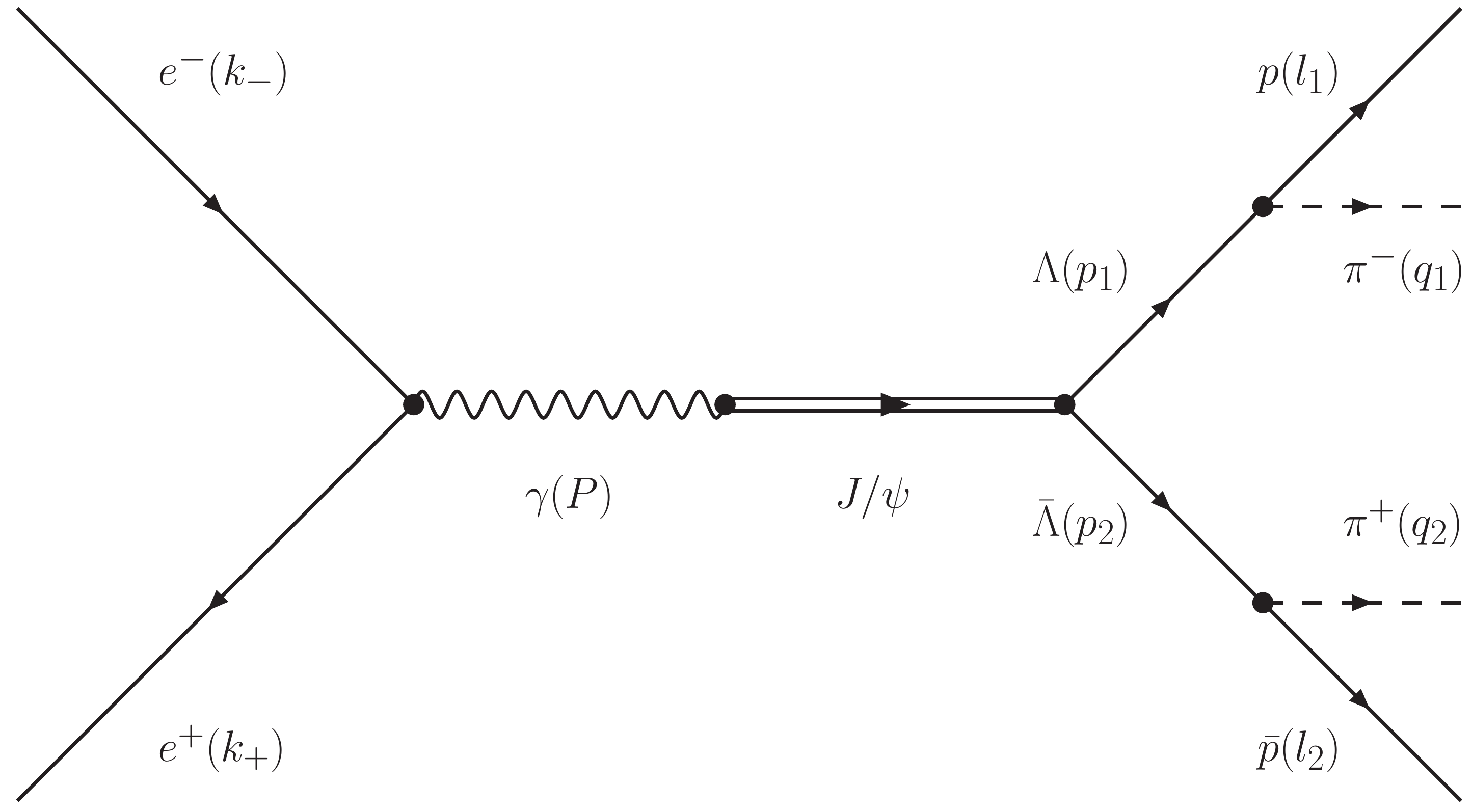}
 \caption[]{\label{fig:diagram} Diagram of the~$e^+e^-\to\jpsi\to[\Lambda\to p\pi^-][\lambar\to
\pbar\pi^+]$ process.}
\end{figure}

In this section we calculate the differential cross section of the process~$e^+e^-\to\jpsi\to[\Lambda\to p\pi^-][\lambar\to
\pbar\pi^+]$ schematically depicted in figure~\ref{fig:diagram}. Then we present various distributions and  asymmetries, which can be built on the basis of this cross section.

\subsection{Kinematics} \label{subsec:kinematics}
\begin{figure}[tb]%
 \centering \includegraphics[width=.8\textwidth]{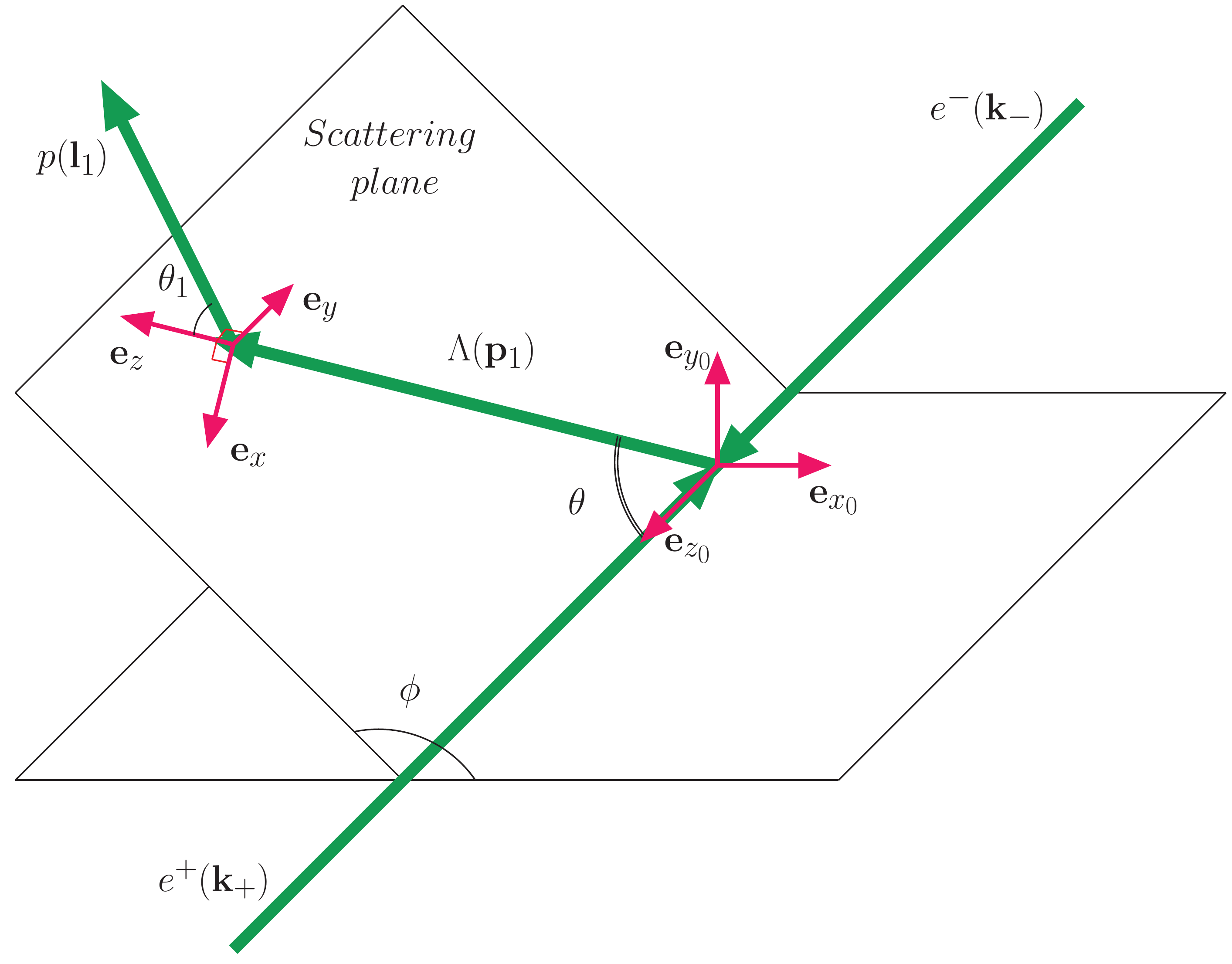}
 \captionsetup{singlelinecheck=off}
 \caption[]{\label{fig:axesChoice}Layout of the lab frame coordinate axes $(\mathbf{e}_{x_0}, \mathbf{e}_{y_0}, \mathbf{e}_{z_0})$ and $\Lambda$ frame coordinate axes $(\mathbf{e}_{x}, \mathbf{e}_{y}, \mathbf{e}_{z})$:
 \begin{eqnarray*}
 \mathbf{e}_z&=&(0,0,1)_{xyz}=(\sin\theta\cos\phi,\sin\theta\sin\phi,\cos\theta)_{x_0y_0z_0},\\
 \mathbf{e}_y&=&(0,1,0)_{xyz}=(\sin\phi,-\cos\phi,0)_{x_0y_0z_0},\\
 \mathbf{e}_x&=&(1,0,0)_{xyz}=(-\cos\theta\cos\phi,-\cos\theta\sin\phi,\sin\theta)_{x_0y_0z_0}.
 \end{eqnarray*}}
\end{figure}
We denote the four-momenta of the particles participating in
the~$e^+e^-\to[\Lambda\to p\pi^-][\lambar\to
\pbar\pi^+]$ process as follows (see figure~\ref{fig:diagram}):
\begin{equation}
 e^{\pm}(k_{\pm}),\quad
 \Lambda(p_1),\quad \lambar(p_2),\quad
 p(l_1),\quad \pbar(l_2),\quad
 \pi^-(q_1),\quad \pi^+(q_2).
\end{equation}
We are going to derive the complete~$5D$ differential distribution of the final-state particles. This problem was solved for the unpolarized electron beam in ref.~\cite{Faldt2017}. We confirm correctness of this result and generalize it to the case of the polarized electron beam adopting the same notation for convenience. We use the following kinematic variables
\begin{equation}\label{kinematics}
  P\equiv k_+ +k_- =p_1+p_2,\quad
  Q\equiv p_1-p_2,\quad
  s\equiv P^2=4m_\Lambda^2-Q^2.
\end{equation}
Here we assume that all particles are on the mass shell: $p_1^2=p_2^2=m_\Lambda^2$, $l_1^2=l_2^2=m_p^2$, $q_1^2=q_2^2=m_\pi^2$, where~$m_\Lambda$ is the $\Lambda$-hyperon mass, $m_p$ is the proton mass, and~$m_\pi$ is the charged pion mass.

As in ref.~\cite{Faldt2017}, we introduce the basis vectors of the right-handed coordinate system in the rest frame of the~$\Lambda$-hyperon ($\Lambda$ frame)
\begin{equation}\label{eq:lambda-frame}
 \mathbf{e}_z = \frac{\mathbf{p}_1}{|\mathbf{p}_1|},\quad
 \mathbf{e}_y = \frac{1}{|\mathbf{p}_1| |\mathbf{k}_-|\sin{\theta}}
 \left(\mathbf{p}_1\times\mathbf{k}_-\right),\quad
 \mathbf{e}_x = \frac{1}{|\mathbf{p}_1| |\mathbf{k}_-|\sin{\theta}}
 \left(\mathbf{p}_1\times\mathbf{k}_-\right)\times
 \frac{\mathbf{p}_1}{|\mathbf{p}_1|},
\end{equation}
where~$\mathbf{k}_-=|\mathbf{k}_-| (0,0,1)$ is the electron's momentum~$3$-vector and
\begin{equation}
    \mathbf{p}_1= |\mathbf{p}_1|\left(\sin\theta\cos\phi, \sin\theta\sin\phi, \cos\theta\right)
\end{equation}
is the~$\Lambda$-hyperon's momentum~$3$-vector in the lab frame, which is the centre-of-mass (\cms) frame (see figure~\ref{fig:axesChoice}); $\theta$ and~$\phi$ are the polar and azimuth angles of~$\Lambda$ in this frame. We will also use~$\theta_1$ and~$\phi_1$ ($\theta_2$ and~$\phi_2$) for the polar and azimuth angles of the proton in the~$\Lambda$ (antiproton in the~$\lambar$) frame and denote the proton's~$3$-momentum and energy in the~$\Lambda$ frame by~$l^{(\Lambda)}_p$ and~$\epsilon_p^{(\Lambda)}$, respectively.

In the~$\Lambda$ frame the proton's~$3$-momentum is~$\mathbf{l}_1=l^{(\Lambda)}_p\hat{\mathbf{l}}_1$, where the unit vector in its direction is
\begin{equation}
    \hat{\mathbf{l}}_1=(\sin\theta_1\cos\phi_1,\sin\theta_1\sin\phi_1,\cos\theta_1),
\end{equation}
as is shown in figure~\ref{fig:axesChoice}.
In the~$\lambar$ frame the antiproton's~$3$-momentum is $\mathbf{l}_2=l^{(\Lambda)}_p\hat{\mathbf{l}}_2$, where the unit vector in its direction reads
\begin{equation}
    \hat{\mathbf{l}}_2=(\sin\theta_2 \cos\phi_2,\sin\theta_2\sin\phi_2,\cos\theta_2).
\end{equation} 

%%% Helicity amplitudes and Lambda form factors %%%
\subsection[Helicity amplitudes and Lambda form factors]
{\boldmath Helicity amplitudes and $\Lambda$ form factors}
\label{subsec:helampl}
We presume that our process is mediated by the $\jpsi$-resonance:
$e^+e^-\to \jpsi \to[\Lambda\to p\pi^-][\lambar\to
\pbar\pi^+]$.
Let us denote the helicities of $\Lambda$, $\lambar$, proton and antiproton by $\lambda_1$, $\lambda_2$, $\lambda'_1$, and $\lambda'_2$ correspondingly keeping $\xi$ for the double helicity of the initial electron.

For the ultrarelativistic electron and positron only the configurations with opposite helicities survive. Therefore the leptonic current reads
\begin{equation}\label{jmue}
  j^\mu_{(e)}\equiv\barr{v}_{-\xi}(k_+) \gamma^\mu
  u_{\xi}(k_-)=\sqrt{s}\left(0, \xi \cos \theta, i, -\xi \sin \theta \right)^\mu,
\end{equation}
where the latter equality is valid in the \cms frame in the axes~$(\mathbf{e}_{x}, \mathbf{e}_{y}, \mathbf{e}_{z})$ shown in figure~\ref{fig:axesChoice};
$\xi=+1$ corresponds to the right-handed electron and $\xi=-1$ to the left-handed one. In this frame and for the same axes the vertex~$\jpsi \to 
\Lambda(p_1,\lambda_1)\lambar(p_2,\lambda_2)$ has the form
\begin{equation}\label{Mlambda}
 \begin{split}
  \Gamma^\mu_\Lambda \left(p_1, p_2\right) & =
   -ie_g {\cal M}^\mu_{\Lambda \lambar}(\lambda_1,\lambda_2) = \\
   & = -ie_g \barr{u}_\Lambda (p_1)\left[{G_M^\psi}\gamma^\mu -
       \frac{2\mlam}{Q^2}\left({G_M^\psi} -{G_E^\psi}\right)Q^\mu\right]v_{\lambar}(p_2)= \\ 
   & = -i e_g 2\sqrt{s}\left(0,
       \lambda_1 G^\psi_M \delta_{\lambda_1,-\lambda_2},
     -\frac{i}{2} G^\psi_M \delta_{\lambda_1,-\lambda_2},
     -\frac{m_\Lambda}{\sqrt{s}} G^\psi_E \delta_{\lambda_1,\lambda_2}
    \right)^\mu.
 \end{split}
\end{equation}
Here the superscript~$^\psi$ is used to avoid confusion between the form factors~${G_E^\psi}$ and~${G_M^\psi}$ and the electromagnetic form factors
of the~$\Lambda$-hyperon. Since the~$\jpsi$ decays into the hyperon pair through three gluons, the~${G_E^\psi}$ and~${G_M^\psi}$ are some effective parameters without universal interpretation.

The invariant amplitude for the~$e^+(k_+,-\xi)e^- (k_-,\xi) \to \jpsi
\to\Lambda(p_1,\lambda_1) \lambar(p_2,\lambda_2)$ process reads
\begin{equation}
 i{\cal M}_{e^+e^-\to\Lambda\lambar} = \frac{i e_g e_\jpsi}{s-m^2_\jpsi+i m_\jpsi \Gamma_\jpsi}
   j_{\mu(e)} {\cal M}^\mu_{\Lambda \lambar}(\lambda_1,\lambda_2),
\end{equation}
where
\begin{equation} \label{JMlambda}
 j_{\mu(e)} {\cal M}^\mu_{\Lambda \lambar}(\lambda_1,\lambda_2)=-2
 m_\Lambda \sqrt{s} G^\psi_E \xi\sin\theta
 \delta_{\lambda_1,\lambda_2} - s G^\psi_M\left(1 + 2\lambda_1\xi \cos\theta \right)\delta_{\lambda_1,-\lambda_2}.
\end{equation}
Here $e_\jpsi$ is the coupling constant of the unpolarized $\jpsi \to e^+e^-$ decay and $e_{g}$ is the
coupling constant of the $\jpsi\to \Lambda
\lambar$ decay defined in ref.~\cite{Faldt2017}
\footnote{There is a misprint in~$e_{g}$ representation in eq.~(A58)  in ref.~\cite{Faldt2017}.}:
\begin{equation}\label{alphas}
 \begin{split}
  \frac{e^2_\jpsi}{4\pi}&\equiv\alpha_\jpsi=\frac{3\Gamma_{\jpsi \to e^+e^-}}{m_\jpsi}, \\
  \frac{e^2_{g}}{4\pi}&\equiv\alpha_{g}=\frac{3\Gamma_{\jpsi \to \Lambda \lambar}}{m_\jpsi\Big(\left|G^\psi_M\right|^2+2m_\Lambda^2/m^2_\jpsi \left|G^\psi_E\right|^2\Big)\sqrt{1-4m_\Lambda^2/m^2_\jpsi}}.
 \end{split}
\end{equation}

One can take into account the electroweak interference between the photon and the~$Z$ boson amplitudes in the~$\jpsi$ production by the substitution~$e_\jpsi \to e^{\xi}_\jpsi$ and correspondingly~$\alpha_\jpsi \to \alpha^{\xi}_\jpsi=(e^{\xi}_\jpsi)^2/(4\pi)$ with
\begin{equation}\label{eq:alphaxi}
 \begin{split}
  e^{\xi}_\jpsi&\approx 
     e_\jpsi\left[1-\xi \frac{-\sinthw+3/8}{4\sinthw(1- \sinthw)}
      \left(\frac{m^2_\jpsi}{m^2_\jpsi-m^2_{Z}+i m_{Z}\Gamma_{Z}}\right) \right], \\
  \alpha^{\xi}_\jpsi & \approx
     \alpha_\jpsi \left[1+\xi\frac{-\sinthw+3/8}{2\sinthw(1-\sinthw)}
       \left(\frac{m_\jpsi}{m_{Z}}\right)^2\right] = \alpha_\jpsi \left(1 + \xi\ahelo \right),
 \end{split}
\end{equation}
where~$\ahelo$ is defined in eq.~\eqref{eq:jpsi_ahelo}. The terms with the factor~$\xi$ in~eq.~\eqref{eq:alphaxi} take into account the difference of the~$Z$ boson coupling with the leptonic current~eq.~\eqref{jmue} for double helicity~$\xi=+1$ and~$\xi=-1$.

The vertex of the transition~$\Lambda (p_1,\lambda_1) \to p
(l_1,\lambda'_1)\pi^-(q_1)$ is described by the invariant
amplitude
\begin{equation} \label{Mpip}
 \begin{split}
   {\cal M}_{(\Lambda)}(\lambda_1,\lambda'_1)&=\barr{u}(l_1)\left[A+B \gamma^5 \right]u(p_1)= \\
   &=\sqrt{2m_\Lambda}e^{i \lambda_1
   \phi_1+i {\pi}(1-2\lambda'_1)/4}
   \sin\left(\frac{\theta_1+\pi(\lambda'_1+\lambda_1)}{2}\right)\times \\
   &\quad\times\left[A \sqrt{\epsilon_p^{(\Lambda)}+m_p}-2\lambda'_1 B
   \sqrt{\epsilon_p^{(\Lambda)}-m_p}\,\right],
 \end{split}
\end{equation}
while the vertex of the transition~$\lambar(p_2,\lambda_2) \to \pbar(l_2,\lambda'_2)\pi^+(q_2)$ is given by the invariant amplitude
\begin{equation} \label{Mantippip}
 \begin{split}
   {\cal M}_{(\lambar)}(\lambda_2,\lambda'_2)&=\barr{v}(p_2)\left[A'+B'
   \gamma^5 \right]v(l_2)= \\
   &=-\sqrt{2m_\Lambda}e^{-i
   \lambda_2 \phi_2+i {\pi}(1-2\lambda'_2)/4}
   \sin\left(\frac{\theta_2+\pi(\lambda'_2-\lambda_2)}{2}\right)\times \\
   &\quad\times\left[A' \sqrt{\epsilon_p^{(\Lambda)}+m_p}-2\lambda'_2 B'
   \sqrt{\epsilon_p^{(\Lambda)}-m_p}\,\right].
 \end{split}
\end{equation}
Following notations of ref.~\cite{Faldt2015}, we use~$R_\Lambda$, $S_\Lambda$ ($\barr{R}_\Lambda$, $\barr{S}_\Lambda$) instead of~$A$, $B$ (and $A'$, $B'$):
\begin{equation}
 \begin{split}
   R_\Lambda &\equiv2
   (l_1 p_1)\left(\left|A\right|^2+\left|B\right|^2\right)+2m_p
   m_\Lambda \left(\left|A\right|^2-\left|B\right|^2\right), \\
   S_\Lambda &\equiv4\operatorname{Re}\left(A^\ast B\right), \\ \barr{R}_\Lambda &\equiv2
   (l_2 p_2)\left(\left|A'\right|^2+\left|B'\right|^2\right)+2m_p
   m_\Lambda \left(\left|A'\right|^2-\left|B'\right|^2\right), \\
   \barr{S}_\Lambda &\equiv4\operatorname{Re}\left({A'}^\ast B'\right).
 \end{split}
\end{equation}

In the final expressions it is convenient to rewrite the form factors through the following dimensionless variables: the ratio~$\alpha$ and the relative phase~$\dphi$ defined as
\begin{equation}\label{eq:ffs1}
 \alpha \equiv
 \frac{s\left|{G_M^\psi}\right|^2 - 4\mlam^2\left|{G_E^\psi}\right|^2}
 {s\left|{G_M^\psi}\right|^2 + 4\mlam^2\left|{G_E^\psi}\right|^2},\quad
 \frac{{G_E^\psi}}{{G_M^\psi}} \equiv
 e^{i\dphi}\frac{\left|{G_E^\psi}\right|}{\left|{G_M^\psi}\right|},
\end{equation}
and to introduce the dimensionless combinations 
\begin{equation}\label{eq:ffs2}
 \alpha_1\equiv -\frac{l^{(\Lambda)}_p m_\Lambda S_\Lambda}{R_\Lambda},\quad\alpha_2\equiv- \frac{l^{(\Lambda)}_p m_\Lambda \barr{S}_\Lambda}{\barr{R}_\Lambda}.
\end{equation}
Note, that~$\alpha$ and~$\dphi$ are functions of the energy invariant $s$. The BESIII Collaboration reported the following values of the form factors at $s=m^2_\jpsi$ in ref.~\cite{kupsc}:
\begin{equation}\label{eq:kupsc}
 \begin{split}
  \dphi &= (42.4 \pm 0.6 \pm 0.5)^{\circ}, \\
  \alpha &= 0.461 \pm 0.006 \pm 0.007, \\
  \alpha_1 &= 0.750 \pm 0.009 \pm 0.004, \\
  \alpha_2 &= -0.758 \pm 0.010 \pm 0.007. 
 \end{split}
\end{equation}

%%% 5D differential distribution %%%
\subsection[5D differential distribution]{$5D$ differential distribution} \label{subsec:5Ddifdistr}
The differential distribution 
\begin{equation}\label{eq:master5d}
 \dd\sigma \propto \wzeta\, \dd(\cos{\theta})\,\dd\Omega_1\,\dd\Omega_2,\quad \dd\Omega_1=\dd\cos\theta_1\dd\phi_1,\quad
 \dd\Omega_2=\dd\cos\theta_2\dd\phi_2,
\end{equation}
depends on the vector~$\zeta$ with~$5$ components
\begin{equation}\label{eq:zeta}
 \zeta \equiv \left(\theta, \theta_1, \phi_1, \theta_2, \phi_2\right).
\end{equation}

The dimensionless quantity~$\wzeta$ is defined via the convolution of the leptonic and hadronic tensors (referred to as the reduced matrix element squared):
\begin{equation}
  \mathcal{W} = \frac{1}{R_\Lambda \barr{R}_\Lambda s \left({s\left\vert G^\psi_M\right\vert^2+4m_\Lambda^2\left\vert
  G^\psi_E\right\vert^2}\right)}L^{\mu \nu}H_{\nu\mu}.
\end{equation}
Detailed calculation of the reduced matrix element squared
\begin{equation}\label{LH}
  \overline{|\mathcal{M}_\mathrm{red}|^2} = L^{\mu\nu}H_{\nu\mu}=a+b\xi
\end{equation}
is presented in the appendix~\ref{app:matrix}. The unpolarized part~$a$ in eq.~\eqref{LH} was previously found in ref.~\cite{Faldt2015}. The explicit expression for~$\wzeta$ reads
\begin{equation}\label{eq:Wdistr}
 \begin{split}
 \wzeta =&
 \,\,\ffi{0}+\alpha\ffi{5}+
 \alpha_1\alpha_2\left(\ffi{1}+\sqral\cosdphi\ffi{2} +\alpha\ffi{6}\right)+\\
 &+\sqral\sindphi\left(\alpha_1\ffi{3}+\alpha_2\ffi{4}\right)+ \\ 
 &+\xi\left[
 (1+\alpha)(\alpha_1\ggi{1}+\alpha_2\ggi{2})+
 \sqral\cosdphi\left(\alpha_1\ggi{3}+\alpha_2\ggi{4}\right)+\right.\\
 &\left.\phantom{\quad\quad}+\sqral\alpha_1\alpha_2\sindphi\ggi{5}\right],
 \end{split}
\end{equation}
where
\begin{equation}
 \begin{split}
 \ffi{0} &=1,\\
 \ffi{1} &=\sin^2\theta\sin\theta_1\sin\theta_2\cos\phi_1\cos\phi_2 + 
 \cos^2\theta\cos\theta_1\cos\theta_2, \\
 \ffi{2} &=\sin\theta\cos\theta\left(
 \sin\theta_1\cos\theta_2\cos\phi_1 +
 \cos\theta_1\sin\theta_2\cos\phi_2\right),\\
 \ffi{3} &=\sin\theta\cos\theta\sin\theta_1\sin\phi_1, \\
 \ffi{4} &=\sin\theta\cos\theta\sin\theta_2\sin\phi_2, \\
 \ffi{5} &=\cos^2\theta, \\
 \ffi{6} &=\cos\theta_1\cos\theta_2 -
 \sin^2\theta\sin\theta_1\sin\theta_2\sin\phi_1\sin\phi_2, \\
%  \\
 \ggi{1} &=\cos\theta\cos\theta_1, \\
 \ggi{2} &=\cos\theta\cos\theta_2, \\
 \ggi{3} &=\sin\theta\sin\theta_1\cos\phi_1, \\
 \ggi{4} &=\sin\theta\sin\theta_2\cos\phi_2, \\
 \ggi{5} &=\sin\theta\left(
 \sin\theta_1\cos\theta_2\sin\phi_1 +
 \cos\theta_1\sin\theta_2\sin\phi_2\right).
 \end{split}
\end{equation}
At~$\xi=0$ these formulas coincide with~eqs.~(6.55)\,--\,(6.56) from ref.~\cite{Faldt2017}. 

Eq.~\eqref{eq:Wdistr} was obtained in the on-shell approximation for the intermediate~$\Lambda$ and~$\lambar$ using effective vertices to describe~$J/\psi$ decay. Corrections to the on-shell propagator approximation are negligible~(${\cal O}( \Gamma_{\Lambda}/m_ {\Lambda})$). Initial state radiative corrections will not alter the ultrarelativistic electron's helicity leaving the angular distributions unchanged. Radiative corrections to the~$\pi^-p$ and~$\pi^+\pbar$ final states are already included into the effective vertices while other radiative corrections with photons or~$Z$ bosons connecting the initial or final state particles are highly suppressed (by~$\alpha^2$) and lack resonance enhancement. Therefore eq.~\eqref{eq:Wdistr} can be safely used for precise measurement of the average polarization~$\pe$ with the accuracy better than~$1\%$.

The quantity~$\mathcal{W}$~\eqref{eq:Wdistr} can be obtained from the helicity amplitudes~\eqref{JMlambda}, \eqref{Mpip}, and~\eqref{Mantippip} as well:
\begin{equation}
  \mathcal{W}=\frac{\frac{1}{2} \sum_{\lambda'_1,\lambda'_2=\pm 1/2}
  \left|\sum_{\lambda_1,\lambda_2=\pm 1/2} j_{\mu(e)} {\cal M}^\mu_{\Lambda
  \lambar}(\lambda_1,\lambda_2)
  {\cal M}_{(\Lambda)}(\lambda_1,\lambda'_1)
  {\cal M}_{(\lambar)}(\lambda_2,\lambda'_2)\right|^2}{R_\Lambda \barr{R}_\Lambda s \left({s\left\vert G^\psi_M\right\vert^2+4m_\Lambda^2\left\vert
  G^\psi_E\right\vert^2}\right)}.
\end{equation}
Further application of the helicity formalism to the baryon-antibaryon pairs produced in collisions of unpolarized electron-positron beams is elaborated in ref.~\cite{heliamp}.

%%% Differential distributions %%%
\subsection{Single-side differential distributions} \label{sec:Difdistr}
Analysis of the full~$5D$ phase space is necessary to measure all parameters involved in eq.~\eqref{eq:Wdistr} if the electron beam is unpolarized. In contrast, the angular distribution of the single-side~$\Lambda\to p\pi^-$ decay contains enough information to disentangle all parameters if the electron beam is polarized. Integration of~eq.~\eqref{eq:Wdistr} over the~$\lambar$ phase space (variables~$\theta_2$ and~$\phi_2$) leads to the~$3D$ differential cross section
\begin{equation}\label{eq:3d-csec}
 \begin{split}
 \frac{\dd\sigma}{\dd\cos{\theta}\,\dd\Omega_1} \propto &
  \,\,1 + \alpha\cos^2{\theta} + 
 \alpha_1 \sqral \sindphi \sin{\theta} \cos{\theta} \sin{\theta_1} \sin{\phi_1}+ \\
 &+\xi\left[\left(1+\alpha\right) \alpha_1 \cos{\theta} \cos{\theta_1} + \alpha_1 \sqral \cosdphi \sin{\theta} \sin{\theta_1} \cos{\phi_1} \right].
 \end{split}
\end{equation}
Note that eq.~\eqref{eq:3d-csec} can be obtained from eq.~\eqref{eq:Wdistr} simply by setting the parameter~$\alpha_2=0$. From the experimental viewpoint, single-side analysis implies inclusive reconstruction of the accompanying~$\lambar$ using the missing mass spectrum. Semi-inclusive single-side reconstruction gives advantage in statistics thanks to both higher reconstruction efficiency and independence from the~$\lambar\to\pbar\pi^+$ branching fraction.

The~$\lambar\to\pbar\pi^-$ decay phase space can be considered completely analogously. The corresponding~$3D$ differential distribution is obtained from~eq.~\eqref{eq:3d-csec} by the substitution $\left(\alpha_1, \phi_1, \theta_1\right) \to \left(\alpha_2, \phi_2, \theta_2\right)$.

%%% Forward-backward asymmetry in the CM frame %%%
\subsubsection[Forward-backward asymmetry in the CM frame]
{Forward-backward asymmetry in the~\cms frame}
Distribution in eq.~\eqref{eq:3d-csec} is expressed in the combined reference frame illustrated in figure~\ref{fig:axesChoice}. The corresponding distribution rewritten via the~\cms frame observables is quite bulky. We place the explicit~$3D$ distribution together with its detailed derivation in appendix~\ref{app:cmsdistrib}. From the practical viewpoint the distribution in the polar angle of the proton in the~\cms frame~$\theta^{(0)}_1$ is particularly interesting.  It is worth noting that there are two values of the proton's energy~$\epsilon_p^{(0)}$ and momentum~$l_p^{(0)}$  for the fixed angles~$\theta_1^{(0)}$ and~$\phi_1^{(0)}$ in the~\cms frame. They correspond to two different configurations of the proton's angles~$\theta_1$ and~$\phi_1$ in the $\Lambda$ frame as we discuss in the appendix, section~\eqref{app:twovaluedness}. The exact distribution can not be represented in elementary functions, but we found the following approximate expression (see details in appendix~\ref{app:cmsdistrib}, eq.~\eqref{sigmap}):
\begin{equation}\label{eq:num_cm}
 \begin{split}
 \frac{\dd\sigma}{\dd\cos\theta^{(0)}_1} \propto &
  \,\,1+\alpha\cos^2\theta_1^{(0)}+ \\
   &+\xi\alpha_1\cos\theta_1^{(0)}\left[0.203\,(1+\alpha)+0.054\sqral\cosdphi \right] +
   {\cal O}\left(\delta^2(s)\right),
 \end{split}
\end{equation}
where
\begin{equation}\label{eq:delta1}
 \delta(s)=\frac{l^{(\Lambda)}_p}{m_p \beta_\Lambda \gamma_\Lambda}=
   \frac{2 m_\Lambda}{\sqrt{s-4m_\Lambda^2}} \frac{l^{(\Lambda)}_p}{m_p}.
\end{equation}
The parameter~$\delta (m_\jpsi^2)\simeq0.1$, so that~eq.~\eqref{eq:num_cm} has a one percent accuracy. More accurate result can be obtained by numerical integration of the exact~$3D$ \cms distribution given in the appendix~\ref{app:cmsdistrib}.

The distribution from~eq.~\eqref{eq:num_cm} is shown in figure~\ref{fig:cmsth}. Electron beam polarization generates the forward-backward asymmetry for protons in the~\cms frame:
\begin{equation}\label{eq:afbcms}
 \mathcal{A}_{\mathrm{FB}}^{(0)} \equiv \frac{
 \int\limits_{0}^{1}{\frac{\dd\sigma}{\dd\cos\theta^{(0)}_1}\dd\cos\theta^{(0)}_1} - 
 \int\limits_{-1}^{0}{\frac{\dd\sigma}{\dd\cos\theta^{(0)}_1}}\dd\cos\theta^{(0)}_1}
 {\int\limits_{0}^{1}{\frac{\dd\sigma}{\dd\cos\theta^{(0)}_1}\dd\cos\theta^{(0)}_1} +
 \int\limits_{-1}^{0}{\frac{\dd\sigma}{\dd\cos\theta^{(0)}_1}\dd\cos\theta^{(0)}_1}}
 \approx 0.11\,\xi.
\end{equation}
Here we use the BESIII results from~eq.~\eqref{eq:kupsc} to obtain the coefficient~$0.11$.

%%%% Left-right asymmetry %%%%
\subsubsection{Left-right asymmetry}
\begin{figure}
 \centering
 \subfloat[]{\label{fig:cmsth}%
 \includegraphics[width=0.45\textwidth]{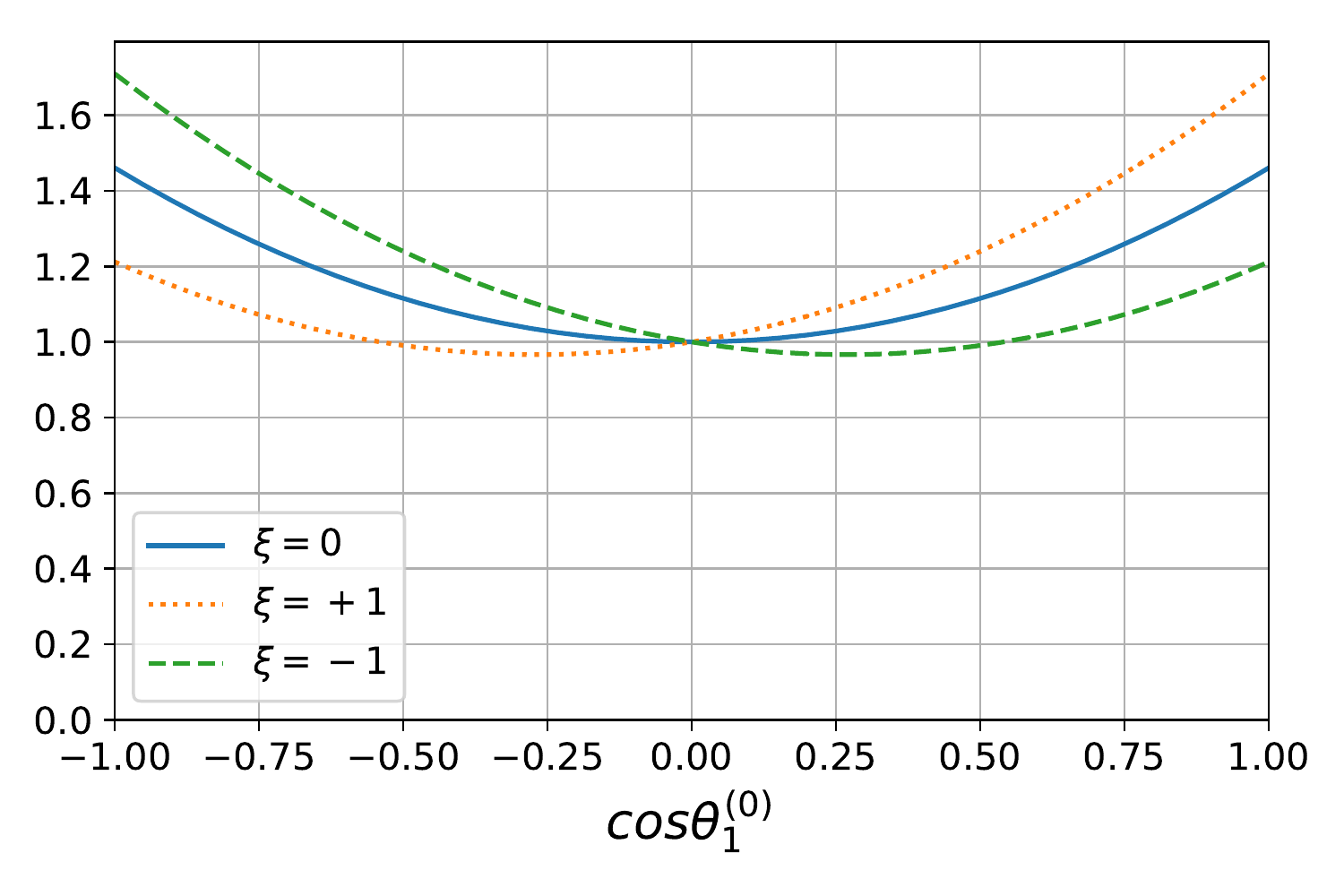}}
 \hfill
 \subfloat[]{\label{fig:phi1}%
 \includegraphics[width=0.45\textwidth]{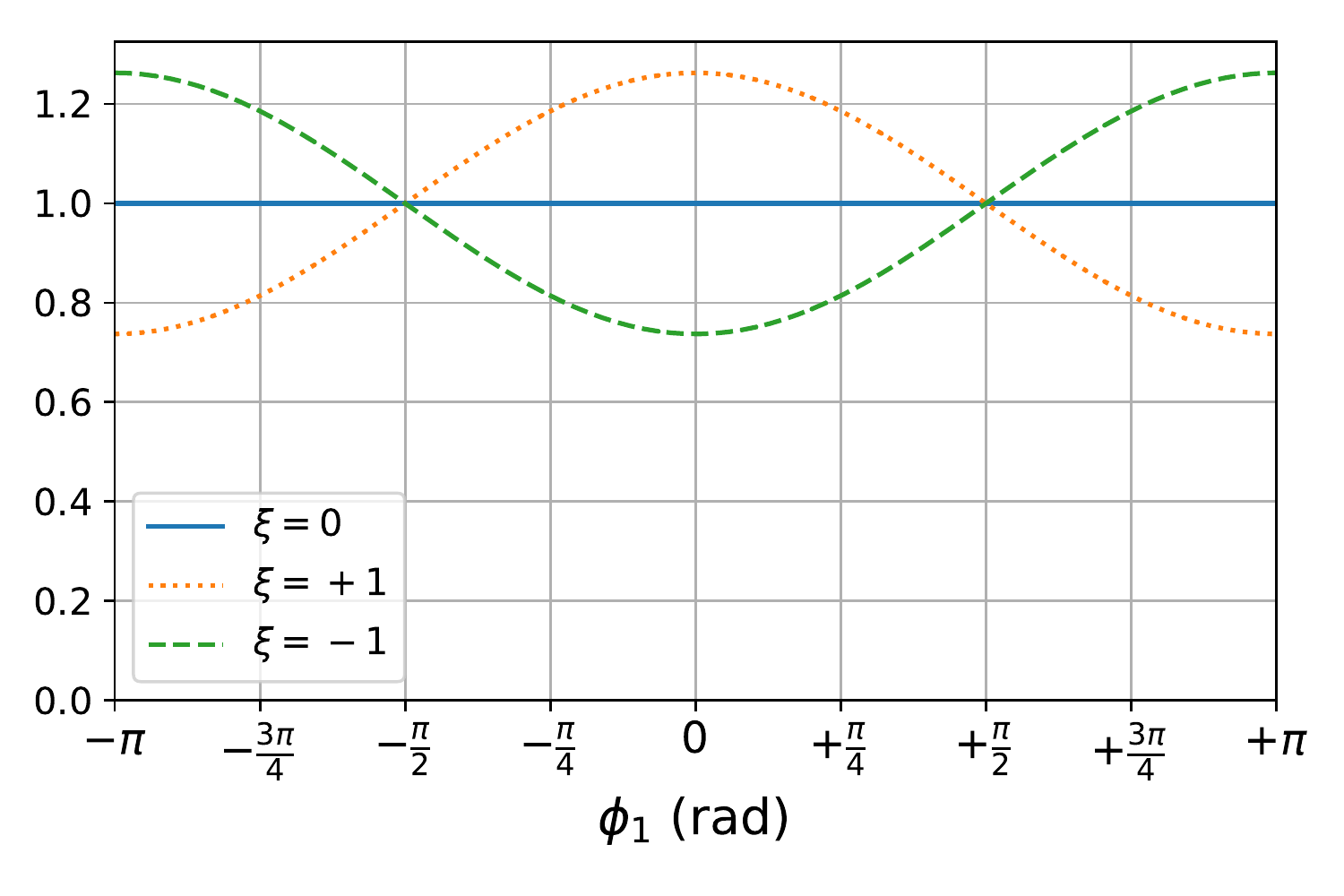}}
 \caption{Distributions in (a) the proton polar angle in the \cms frame and (b) the proton azimuth angle in the~$\Lambda$ frame for the~$e^+e^-\to\jpsi\to[\Lambda\to p\pi^-]\lambar$ process. Solid blue lines correspond to the unpolarized electron beam ($\xi=0$), dotted orange lines correspond to the beam of electrons with double helicity~$\xi=+1$, and the dashed green lines correspond to the beam of electrons with double helicity~$\xi=-1$.}
 \label{fig:1d_distr}
\end{figure}

The distribution in the azimuth angle~$\phi_1$,
\begin{equation}\label{eq:dsigmadphi1}
 \frac{\dd\sigma}{\dd\phi_1} \propto 1 + \frac{\alpha}{3} + \xi\frac{\pi^2}{16}\alpha_1\sqral\cosdphi\cos{\phi_1},
\end{equation}
is obtained by integration of~eq.~\eqref{eq:3d-csec} over~$\costh$ and~$\costh_1$. It is sensitive to the electron beam polarization, as is illustrated in figure~\ref{fig:phi1}. Further reduction leads us to the integral observables:
\begin{equation}
 \begin{split}
 \sigl &\equiv \int\limits_{-\pi/2}^{\pi/2} \frac{\dd\sigma}{\dd\phi_1} \dd\phi_1 \propto 1 + \frac{\alpha}{3} +\xi\frac{\pi}{8}\alpha_1\sqral\cosdphi,\\
 \sigr &\equiv \int\limits_{\pi/2}^{3\pi/2} \frac{\dd\sigma}{\dd\phi_1} \dd\phi_1 \propto 1 + \frac{\alpha}{3} -\xi\frac{\pi}{8}\alpha_1\sqral\cosdphi,
 \end{split}
\end{equation}
and the azimuthal left-right asymmetry
\begin{equation}\label{eq:alr}
 \alr \equiv \frac{\sigl - \sigr}{\sigl + \sigr} = \xi \frac{3\pi}{8}\frac{\sqral}{\alpha + 3}\alpha_1\cosdphi\approx 0.17\,\xi.
\end{equation}
The coefficient~$0.17$ in~eq.~\eqref{eq:alr} is obtained using the experimental data from~eq.~\eqref{eq:kupsc}.

%%%% Forward-backward asymmetry %%%%
\subsubsection[2D forward-backward asymmetry]{$2D$ forward-backward asymmetry}\label{sec:afb}
\begin{figure}
 \centering
 \subfloat[]{\label{fig:bf_xi0}%
 \includegraphics[width=0.45\textwidth]{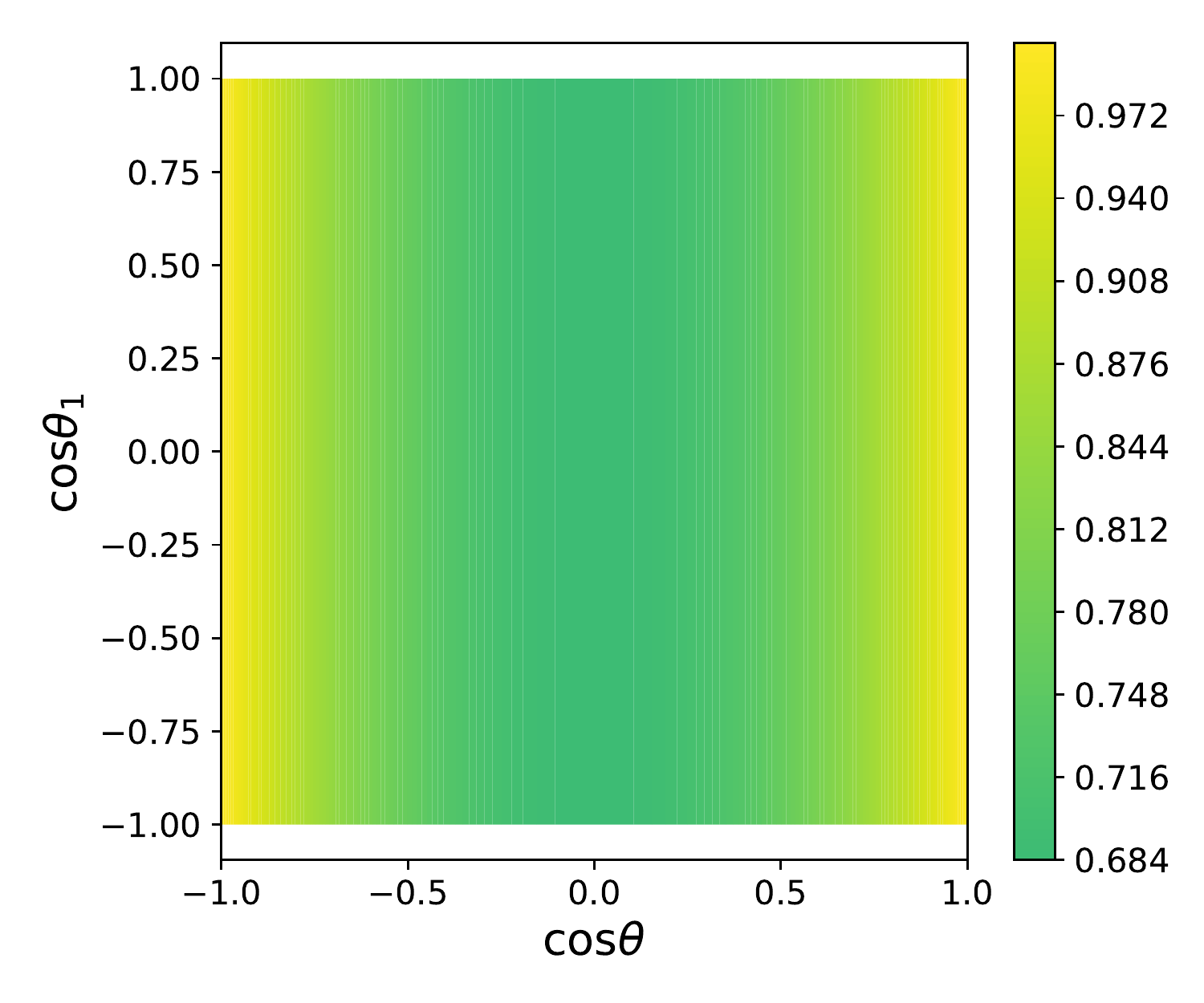}}
 \hfill
 \subfloat[]{\label{fig:bf_xi1}%
 \includegraphics[width=0.45\textwidth]{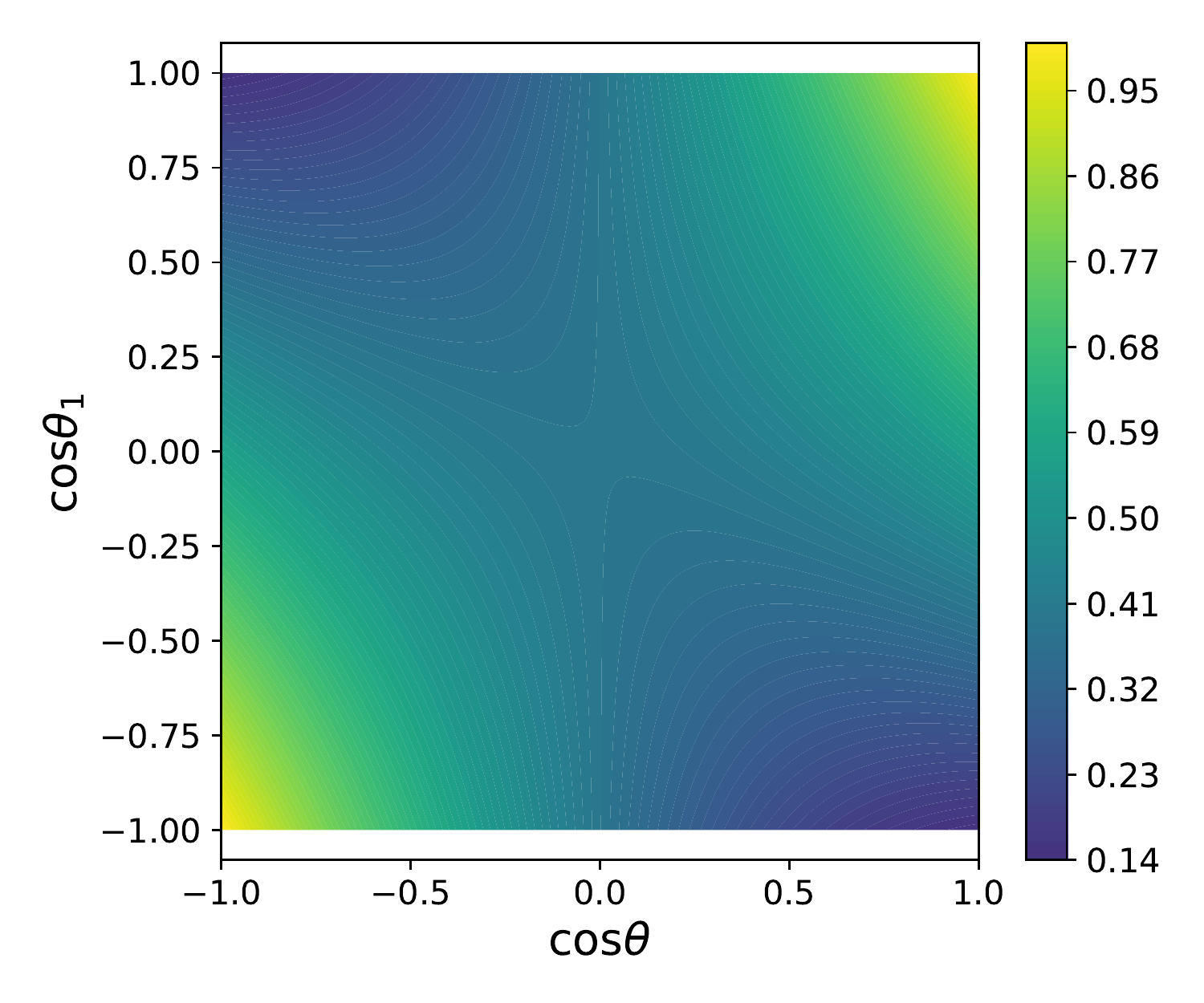}}
 \caption{Differential distribution in the polar angles~$\theta$ and~$\theta_1$ from eq.~\eqref{eq:2d-csec}. Left: unpolarized beam, $\xi=0$. Right: polarized beam with $\xi=+1$. Color scale is the same for the both plots.}
 \label{fig:theta_distr}
\end{figure}

Integration of eq.~\eqref{eq:3d-csec} over the azimuth angle~$\phi_1$ leads to the~$2D$ distribution in the polar angles~$\theta$ and~$\theta_1$,
\begin{equation}\label{eq:2d-csec}
 \frac{\dd\sigma}{\dd\cos{\theta}\,\dd\cos{\theta_1}} \propto 1 + \alpha\cos^2{\theta} + \xi\left(1+\alpha\right)\alpha_1\cos{\theta}\cos{\theta_1}.
\end{equation}
Beam polarization makes the~$\costh$ and~$\costh_1$ distributions correlated as is illustrated in figure~\ref{fig:theta_distr}. This correlation has a simple interpretation in the lab frame (see eq.~\eqref{eq:num_cm}): protons tend to fly along the electron beam polarization (and antiprotons tend to fly in the opposite direction since~$\alpha_2$ is negative).

Integral observables corresponding to the distribution in eq.~\eqref{eq:2d-csec}:
\begin{equation}
 \begin{split}
   \sfwd \equiv \int\limits_{\cos{\theta}\cos{\theta_1} > 0}
   \frac{\dd\sigma}{\dd\cos{\theta}\,\dd\cos{\theta_1}}  \dd\cos{\theta}\,\dd\cos{\theta_1},\\
   \sbwd \equiv \int\limits_{\cos{\theta}\cos{\theta_1} < 0}
   \frac{\dd\sigma}{\dd\cos{\theta}\,\dd\cos{\theta_1}} \dd\cos{\theta}\,\dd\cos{\theta_1},
 \end{split}
\end{equation}
lead to the forward-backward asymmetry
\begin{equation}\label{eq:afb}
 \afb = \frac{\sfwd - \sbwd}{\sfwd + \sbwd} = \xi\frac{3\alpha_1}{4}\frac{\alpha+1}{\alpha+3}
 \approx 0.24\,\xi.
\end{equation}

%%%%%%%%%%%%%%%%%%%%%%%%%%%
%%%% Feasibility study %%%%
%%%%%%%%%%%%%%%%%%%%%%%%%%%
\section{Feasibility study}\label{sec:feas}
The results obtained in section~\ref{sec:matrix_element} are supposed to be used in the data analysis at a future SCT experiment. The developed formalism allows one to precisely measure the parameters~$\alpha$, $\dphi$, $\alpha_1$ and~$\alpha_2$ together with the average electron beam polarization~$\pe$. Choice of the specific analysis strategy is a trade-off between statistical and systematic uncertainties. The complete~$5D$ phase space analysis employing eq.~\eqref{eq:Wdistr} provides the best statistical precision but leads to a difficult~$5D$ analysis of the detection efficiency. The opposite extreme is a counting-experiment measurement of the average polarization~$\pe$ employing any of the integral asymmetries from~eqs.~\eqref{eq:afbcms}, \eqref{eq:alr}, and~\eqref{eq:afb}. 

The expected statistical precision for the parameters~$\alpha$, $\dphi$, $\alpha_1$ and~$\alpha_2$ and the average electron beam polarization~$\pe$ is addressed in this section. Three measurement schemes are considered: the full~$5D$ fit employing~eq.~\eqref{eq:Wdistr}, the single-side~$3D$ fit employing~eq.~\eqref{eq:3d-csec}, and the counting experiments. The objectives are to assess the statistical sensitivity to the average polarization~$\pe$ and to evaluate what effect the electron beam polarization has on the precision of measurement of the other parameters.

The following procedure is used in the feasibility study:
\begin{enumerate}
 \item The $\jpsi\to [\Lambda\to p\pi^-][\lambar\to\pbar\pi^+]$ events with uniform (phase-space) momentum distribution are generated with the~\texttt{EvtGen} generator (see ref.~\cite{evtgen}) embedded in the SCT software framework \texttt{AURORA} (see ref.~\cite{aurora}).
 \item The signal events are obtained with accept-reject procedure employing the~$\wzeta$ distribution~from~eq.~\eqref{eq:Wdistr} as probability density.
 \item Simple selection criteria are imposed on the events: transverse momentum of the final-state particles is to be higher than~$60~\mathrm{MeV}/c$ and the angle between the beam direction and the particle momentum is to be larger than~$10^{\circ}$.
 \item The unbinned maximum likelihood fit is performed to obtain the parameters defined in~eqs.~\eqref{eq:ffs1} and~\eqref{eq:ffs2}. The following likelihood function is used:
 \begin{equation}\label{eq:lhfcn}
 -2\ln\mathcal{L} = -2\sum\limits_{i=1}^{N}\ln{\wzetai} + 2N\ln{\sum\limits_{j=1}^{M}\wzetaj}.
 \end{equation}
 Here~${\cal W}$ is the decay probability density under study, $\zeta$ is defined in eq.~\eqref{eq:zeta}, ~$\tilde{\zeta}_j$ correspond to the phase-space events,~$\zeta_i$ correspond to the events weighted with~${\cal W}$,~and the condition $M\gg N$ is respected. Minimization of the likelihood function in~eq.~\eqref{eq:lhfcn} is performed with the \texttt{MINUIT} algorithm (see ref.~\cite{minuit}) via the \texttt{iminuit} (see ref.~\cite{iminuit}) \texttt{python} interface.
\end{enumerate}

The expected annual signal yield of $\jpsi\to [\Lambda\to p\pi^-][\lambar\to\pbar\pi^+]$ events at an SCT factory is
\begin{equation}\label{eq:nsig}
 N_{\mathrm{sig}} = N_\jpsi\times \brjpsilamlam \times \left(\brlamppi\right)^2 \times \epsdet
 = 0.8\times 10^9\epsdet,
\end{equation}
where~$\epsdet$ is the detection efficiency, $N_\jpsi=10^{12}$ is the expected number of~$\jpsi$ states produced at the SCT factory during one data taking season, and the branching fractions (see ref.~\cite{pdg})
\begin{equation}
 \brjpsilamlam = (1.89 \pm 0.09)\times 10^{-3},\quad
 \brlamppi = (63.9 \pm 0.5)\%.
\end{equation}

\subsection{Estimates for the statistical precision}\label{sec:feas-stat}
The maximum likelihood fit procedure is applied to the following analysis schemes:
\begin{enumerate}
 \item Full reconstruction $5D$ fit with unpolarized beams and~$\wzeta$ defined in~eq.~\eqref{eq:Wdistr}.
 \item Full reconstruction $5D$ fit with the average beam polarization~$\pe=0.8$ and~$\wzeta$ defined in~eq.~\eqref{eq:Wdistr}.
 \item Single-side reconstruction $3D$ fit with the average beam polarization~$\pe=0.8$ and~$\wzeta$ defined in~eq.~\eqref{eq:3d-csec}.
\end{enumerate}

The first scheme implies four free parameters defined in~eqs.~\eqref{eq:ffs1} and~\eqref{eq:ffs2}. The other two schemes have two additional free parameters: the average polarizations of the electron beam corresponding to the data sets with the right-handed and the left-handed electrons, respectively. The statistical precision obtained for the expected signal yield at the SCT factory is shown in table~\ref{tab:full-fit}.

\begin{table}[ht]
\renewcommand{\arraystretch}{1.5}
\centering
 \begin{tabular}{|l|cccc|} \hline
 \multirow{2}{*}{Analysis scheme} & \multicolumn{4}{c|}{SCT one-year $\sigma$ ($10^{-4}$)} \\
 \cline{2-5}
 & $\pe$ & $\alpha$ & $\dphi\ (\mathrm{rad})$ & $\alpha_i$ \\ \hline
 Full reconstruction $5D$,~$\pe=0$ & Fixed & $1.5$ & $3.1$ & $2.8$ \\
 Full reconstruction $5D$,~$\pe=0.8$ & $1.3$ & $1.2$ & $1.6$ & $0.9$ \\
 Single-side reconstruction $3D$,~$\pe=0.8$ & $4.3$ & $1.2$ & $2.4$ & $3.4$ \\
 \hline
 \end{tabular}
 \caption{The expected one-year statistical precision for the parameters defined in~eqs.~\eqref{eq:ffs1} and~\eqref{eq:ffs2} and the average polarization~$\pe$ obtained through different experimental schemes.}
 \label{tab:full-fit}
\end{table}

The first result obtained is that the~$5D$ analysis at the SCT factory with the polarized electron beam provides the statistical precision level of order of~$10^{-4}$ for all parameters. In particular, this precision for the average beam polarization is good enough for the weak mixing angle measurement. The single-side analysis provides about three times worse precision of the average polarization monitoring, but still satisfies the requirements for the weak mixing angle measurement. 

Electron beam polarization leads to considerable improvement in statistical precision of the parameter measurement. In particular, the precision of the phase difference~$\dphi$ is improved by a factor of two. The most significant improvement of about three times occurs for the precision of the~$\Lambda$ and~$\lambar$ decay parameters~$\alpha_1$ and~$\alpha_2$. We will explain this fact in the next section. The single-side analysis with the polarized electron beam (third row in table~\ref{tab:full-fit}) provides statistical precision similar to that of the full reconstruction analysis with the unpolarized beams, but with the potentially better control of systematic uncertainties.

\subsection{Sensitivity to $CP$ violation in the $\Lambda\to p\pi^-$ decay}
\begin{figure}
 \centering
 \subfloat[]{\label{fig:corr-upol}%
 \includegraphics[width=0.48\textwidth]{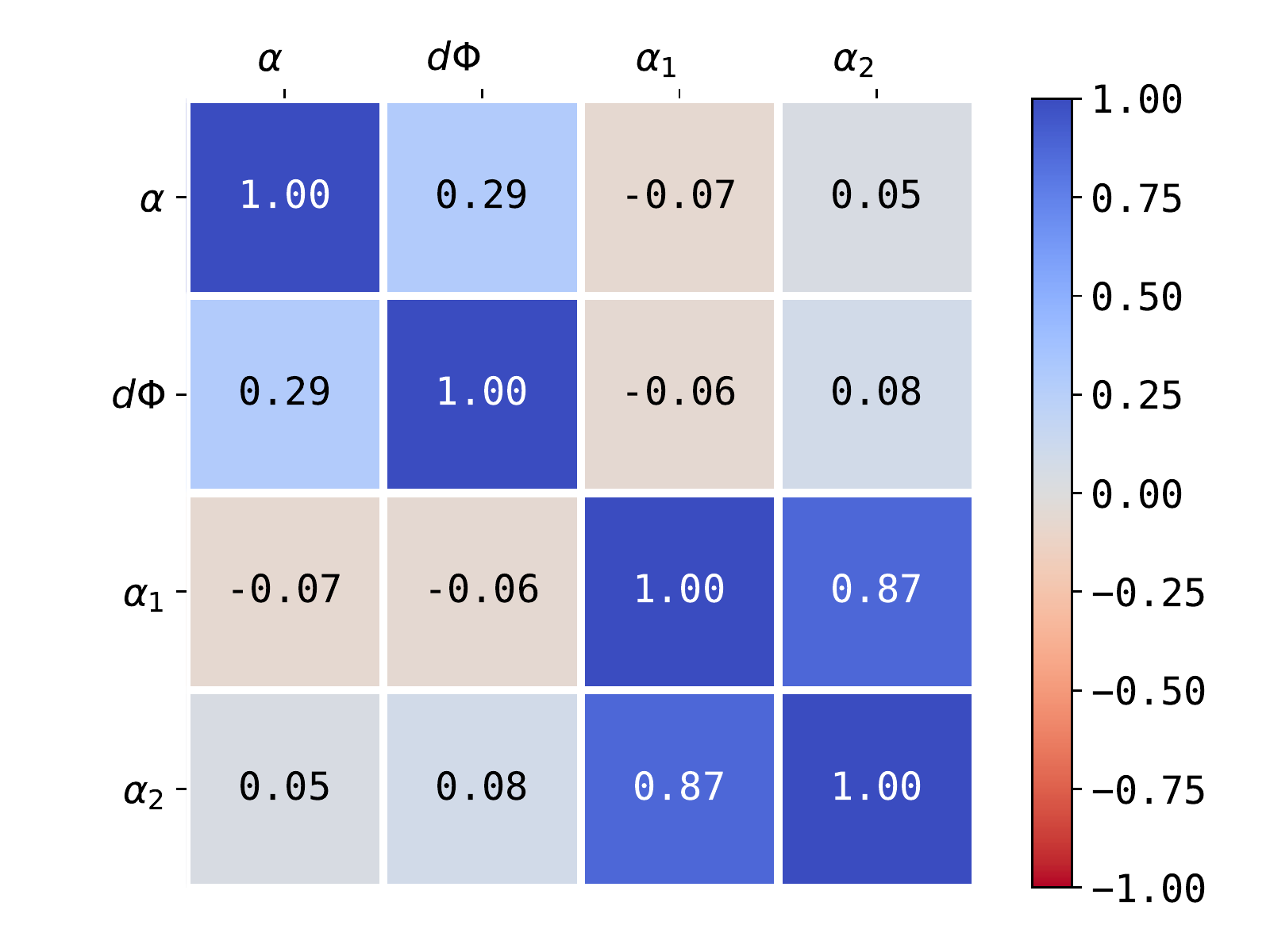}}
 \hfill
 \subfloat[]{\label{fig:corr-xi0_8}%
 \includegraphics[width=0.48\textwidth]{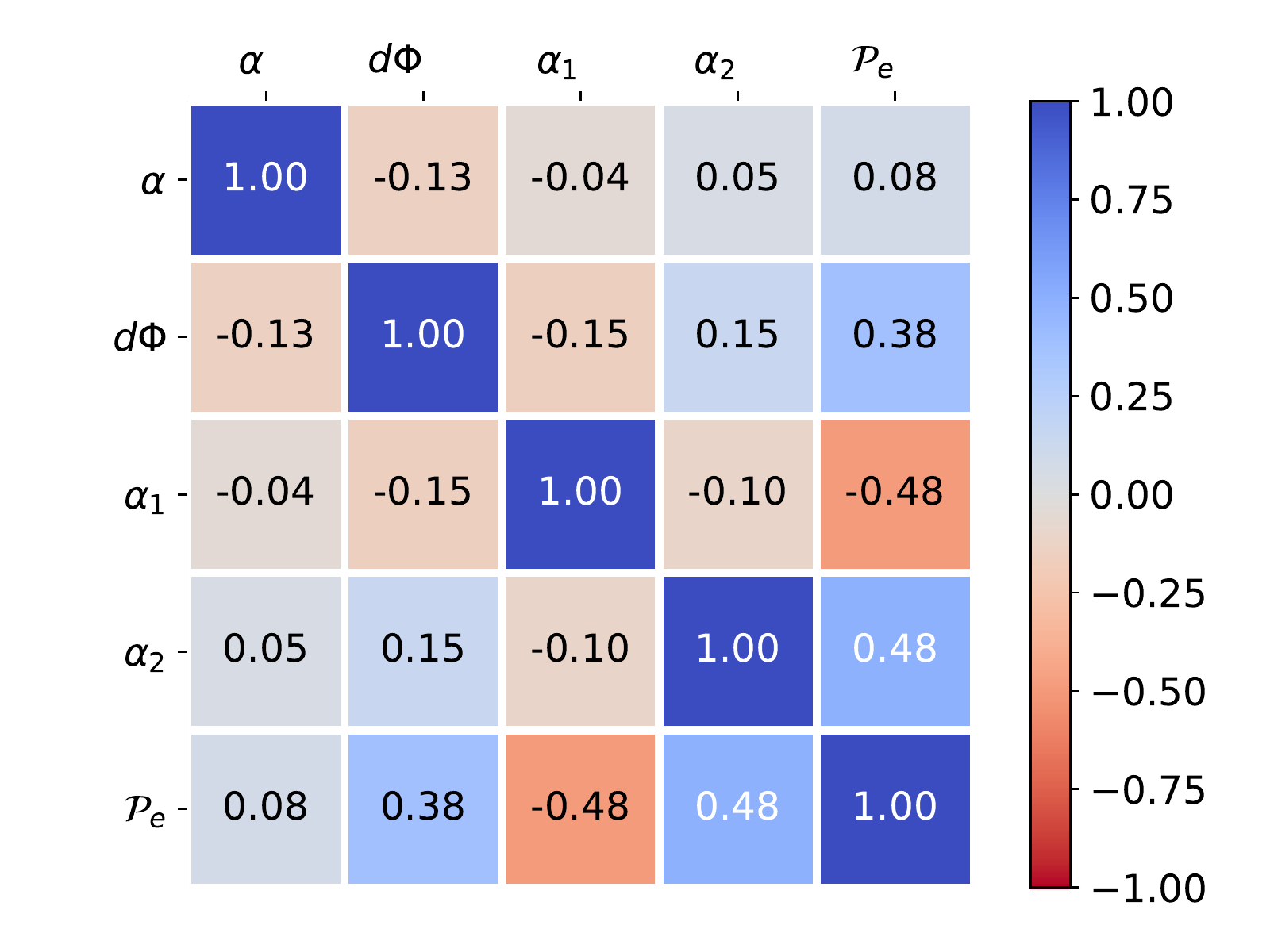}}
 \caption{Correlation matrices for the fit schemes $1$ (a) and $2$ (b). Note the strong correlation of~$\alpha_1$ and~$\alpha_2$ in scheme~$1$ (unpolarized beam) and absence of this correlation in scheme~$2$ (average electron beam polarization~$\pe=0.8$).}
 \label{fig:corr-mtx}
\end{figure}

$CP$ symmetry implies~$\alpha_1 = -\alpha_2$, so a deviation of the sum~$(\alpha_1+\alpha_2)$ from zero would manifest $CP$ symmetry breaking. The standard model predicts a very small value for the $CP$ asymmetry (see ref.~\cite{cpv_in_lambda})
\begin{equation}\label{eq:acp-lam}
 \left|{\cal A}_{CP}\right| \equiv 
 \left|\frac{\alpha_1+\alpha_2}{\alpha_1-\alpha_2}\right| < 5 \times 10^{-5}.
\end{equation}
Electron beam polarization improves precision in the measurement of~$\alpha_1$ and~$\alpha_2$ and therefore enhances sensitivity to $CP$ violation in the~$\Lambda\to p\pi^-$ decay. Moreover, the correlation coefficient between the parameters~$\alpha_1$ and~$\alpha_2$ in the case of the unpolarized beam (scheme~$1$) is close to~$+1$ as shown in figure~\ref{fig:corr-upol}. So the fit is more sensitive to the difference~$\alpha_1-\alpha_2$ than to the $CP$ violating sum~$\alpha_1+\alpha_2$ (and~${\cal A}_{CP}$).
This conclusion is confirmed by the fit with explicit change of variables:
\begin{equation}
 s \equiv \frac{1}{2}\left(\alpha_1 + \alpha_2\right),\quad
 d \equiv \frac{1}{2}\left(\alpha_1 - \alpha_2\right).
\end{equation}
The fit precision of~$d$ is about three times better than that of~$s$.

The correlation matrix for the scheme~$2$ fit is shown in figure~\ref{fig:corr-xi0_8}. Note that the correlation between the four parameters decreases. The most significant effect is for the correlation between~$\alpha_1$ and~$\alpha_2$. It can be interpreted as follows: the~$\Lambda$ and~$\lambar$ sides become less correlated because each side carries enough information to disentangle the decay dynamics, thanks to the beam polarization. It explains the significant improvement of statistical precision for~$\alpha_1$ and~$\alpha_2$ in the scheme~$2$ indicated in table~\ref{tab:full-fit}.

%%%%% Reduced phase space %%%%%
\subsection{Counting experiments}
Asymmetries in~eqs.~\eqref{eq:afbcms}, \eqref{eq:alr}, and~\eqref{eq:afb} can be used to measure the average beam polarization with a counting experiment. The asymmetries are proportional to the average polarization~$\pe$ with some coefficient~$\eta\sim 0.1$ dependent on the parameters. The parameters should be measured independently. One option is to use the data set with the unpolarized electron beam to measure the parameters (scheme~$1$ from section~\ref{sec:feas-stat}). Statistical precision for the average beam polarization in this case reads
\begin{equation}
 \sigma(\pe) = \frac{1}{\eta}\left(\frac{1}{\sqrt{N_0}} \oplus \pe\sigma(\eta)\right),
\end{equation}
where~$N_0\approx N_{\mathrm{sig}}/3$ (see eq.~\eqref{eq:nsig}) is the number of events in the sample with a certain polarization. Statistical precision~$\sigma(\pe)$ of~$0.1\%$ is reachable with this approach.

%%%%%%%%%%%%%%%%%%%%%%%%%%%%%%%%%
%%%%% Collider requirements %%%%%
%%%%%%%%%%%%%%%%%%%%%%%%%%%%%%%%%
\section{Experimental effects}\label{sec:systematics}
New precision frontiers pose new challenges in data analysis and reveal subtle effects that used to be safely neglected. This section presents a brief review of experimental effects foreseeing the weak mixing angle measurement at the Super Charm-Tau factory with data-driven monitoring of the average beam polarization.

\subsection{Luminosity monitoring}
The asymmetry in~$\ahel$~eq.~\eqref{eq:jpsi_ahel} is expressed in terms of the cross sections~$\sigma_{\scriptscriptstyle \pe}$ and~$\sigma_{\scriptscriptstyle -\pe}$. The actual observables are event counts~$N_{\scriptscriptstyle \pe}$ and~$N_{\scriptscriptstyle -\pe}$ corresponding to the data sets collected with the opposite average beam polarizations:
\begin{equation}
 \sigma_{\scriptscriptstyle \pe} = \frac{N_{\scriptscriptstyle \pe}}{\mathcal{L}_{\scriptscriptstyle \pe}\varepsilon_{\mathrm{eff}}},\quad
 \sigma_{\scriptscriptstyle -\pe} = \frac{N_{\scriptscriptstyle -\pe}}{\mathcal{L}_{\scriptscriptstyle -\pe}\varepsilon_{\mathrm{eff}}}.
\end{equation}
It is likely safe to assume the same detection efficiency~$\varepsilon_{\mathrm{eff}}$ for both data sets. In contrast, the luminosity integrals~$\mathcal{L}_{\scriptscriptstyle \pe}$ and~$\mathcal{L}_{\scriptscriptstyle -\pe}$ can not be expected to be equal precisely enough in general and should be measured. The statistical uncertainty of the measured luminosity must be better than~$1/\sqrt{N_{\scriptscriptstyle \pe,-\pe}}\sim 10^{-6}$. The uncertainty in the measured luminosity translates into the uncertainty in the cross sections.

A conventional way of luminosity monitoring is analysis of the Bhabha scattering events. Cross section of the detectable Bhabha scattering events is determined by hermeticity of the detector. The minimal polar angle~$\theta=10^{\circ}$, which is an optimistic assumption, leads to the Bhabha scattering cross section of about~$1\times 10^{-30}~\mathrm{cm}^2$, which is three times less than the~$\jpsi$ cross section from~eq.~\eqref{eq:jpsicsec}. It means that the luminosity monitoring will give a sizeable contribution to the~$\ahel$ measurement and potentially limit the achievable precision. A dedicated detector counting collinear Bhabha scattering events at low angles would allow one to get large enough statistics and break this limitation. This detector should be sufficiently fast to measure luminosity bunch-by-bunch and be able to cope with the high radiation load.

A multiplicative systematic uncertainty shared among~$\mathcal{L}_{\scriptscriptstyle \pe}$ and~$\mathcal{L}_{\scriptscriptstyle -\pe}$ does not affect the observable asymmetry. The requirement of the shared systematic uncertainty can be met only if both data sets are collected simultaneously. It can be done since about~$500$ electron bunches are to circulate in the collider ring at the same time (see ref.~\cite{sct}), and electron polarization can be set in a bunch-by-bunch regime.

To conclude this discussion consider common additive background:
\begin{equation}
 N_{\scriptscriptstyle \pe}^{\prime} = N_{\scriptscriptstyle \pe} + N_{\mathrm{bkg}},\quad
 N_{\scriptscriptstyle -\pe}^{\prime} = N_{\scriptscriptstyle -\pe} + N_{\mathrm{bkg}}.
\end{equation}
This kind of background should be suppressed up to the level of~$\ahel$ statistical precision: $N_{\mathrm{bkg}} / N_{\scriptscriptstyle \pe} \lesssim 10^{-3}$ (not up to~$10^{-6}$ as can be naively assumed), that looks like a feasible constraint.

A very different, but conceivable, option to consider is to provide very stable bunch current and bunch crossing conditions to guarantee that the integrated luminosities $\mathcal{L}_{\scriptscriptstyle \pe}$ and~$\mathcal{L}_{\scriptscriptstyle -\pe}$ are equal with precision better than~$10^{-6}$.

\subsection{Spin rotation in the detector field}
The $\Lambda$ baryon has the lifetime~$\tau\approx\ 2.6\times 10^{-10}~\mathrm{s}$ and the corresponding decay length of several centimeters. Its spin rotates around the detector magnetic field~$\mathbf{B}$ with the angular velocity (see ref.~\cite{Berestetskii})
\begin{equation}
 \bm{\omega} = -\frac{2\mu_\Lambda + 2\mu'_\Lambda(\gamma_\Lambda - 1)}{\hbar\gamma_\Lambda}\mathbf{B}+\frac{2\mu'_\Lambda\gamma_\Lambda}{\hbar(\gamma_\Lambda + 1)} (\bm{\beta}_\Lambda\mathbf{B})\bm{\beta}_\Lambda,
\end{equation}
where $\mu_\Lambda$ and $\mu'_\Lambda$ are the total and the anomalous magnetic moments of the $\Lambda$ baryon, respectively, and
$\gamma_\Lambda=\sqrt{s}/(2m_\Lambda)$ is its Lorentz factor with the velocity $\bm{\beta}_\Lambda$, $\beta_\Lambda=\sqrt{1-4m_\Lambda^2/s}$.

Electric charge of $\Lambda$ is zero, thus $\mu_\Lambda=\mu'_\Lambda$ and
\begin{equation}
 \bm{\omega} = \frac{-2\mu_\Lambda}{\hbar} \left(\mathbf{B}-
 \frac{\gamma_\Lambda}{\gamma_\Lambda+1}(\bm{\beta}_\Lambda\mathbf{B})\bm{\beta}_\Lambda\right).
\end{equation}
The rotation angle is maximal when the magnetic field and the baryon's velocity are perpendicular,
\begin{equation}
\theta_{\mathrm{max}}=\omega_{\mathrm{max}}\tau^{(0)}=\frac{2|\mu_\Lambda|}{\hbar}B\gamma_\Lambda \tau.
\end{equation}

The magnetic moment of $\Lambda$ is $\mu_\Lambda =(-0.613\pm 0.004)\,\mu_N$, where $\mu_N=|e|\hbar/(2m_p c)$ is the nuclear magneton and $\gamma_\Lambda \approx 1.39$ for $s=m_\jpsi^2$. Therefore, the magnetic field of~$B=1.5$~T leads to $\theta_{\mathrm{max}} \approx 32~\mathrm{mrad}$.

This rotation affects the observable angular distribution and probably should be taken into account to achieve sub-percent precision level for~$\sinthw$. Fortunately, an event-by-event correction can be applied using the measured flight length of~$\Lambda$, which imposes certain requirements on tracking system and vertex resolution. See discussion of this effect in ref.~\cite{detfield}.

\subsection{Deflection of particles in the beam}
A large number of particles in each beam and small bunch size lead to a significant magnetic field generated by the bunch. The BINP SCT project has the following parameters (see ref.~\cite{sct}): single bunch current of~$4.2~\mathrm{mA}$ and beam size of $0.178~\mathrm{\mu m}\times 17.8~\mathrm{\mu m}\times 10~\mathrm{mm}$. They give the magnetic field of about~$0.01~\mathrm{T}$ at the surface of the flat beam. $\Lambda$ spin rotation in this field is negligible, but such a bunch field can disturb the Bhabha distribution and introduce a bias to the luminosity measurement, as is discussed in detail in ref.~\cite{beamfield}.

\subsection{Finite beams crossing angle and natural polarization}
Differential cross sections presented in section~\ref{sec:Difdistr} are derived under the assumption of collinear electron-positron bunch crossing. The crab-waist beam collision scheme implies the crossing angle of about~$60~\mathrm{mrad}$ leading formally to a different setup. The helicity state of an ultrarelativistic electron is invariant under weak boosts. Therefore our formulas remain applicable.

More attention should be paid to a possible polarization of positrons. It is known that there is a natural polarization of particles in collider rings that is parallel or antiparallel to the magnetic field, i.e. transverse to the plane of the ring~(see ref.~\cite{st}). Our calculations imply unpolarized positrons. Nevertheless, such a  transverse positron polarization (we denote it as~$\vec{\zeta}_{\bot}$ in Appendix \ref{subsec:lephadtens}) does not change our results. Indeed, the terms proportional to~$\vec{\zeta}_{\bot}$ cancel in the leptonic tensor as is shown in eq.~\eqref{Appleptonictensorgen}. Longitudinal positron polarization~$\zeta_{\shortparallel}$, that can potentially change our results, can also appear due to non-linear beam dynamics. A way to eliminate all possible effects related to positron polarization is to foresee a device depolarizing the positron beams.

%%%%%%%%%%%%%%%%%%%%%%%
%%%%% Conclusions %%%%%
%%%%%%%%%%%%%%%%%%%%%%%
\section{Conclusions}\label{sec:conc}
Measurement of the weak mixing angle at a Super Charm-Tau factory experiment would provide a unique probe of the neutral weak coupling of the charm quark at low, relative to~$m_Z$, momentum transfer. The expected statistical precision of the measurement, $\delta\left(\sinthw\right)/\sinthw\approx 0.3\%$, approaches the most precise at the moment results of LEP and SLD (see ref.~\cite{lepslc}). 

This weak mixing angle measurement is a challenging experiment that should be carefully planned ahead. Longitudinal polarization of the electron beam is a necessary condition for this experiment. The data with positive and negative beam polarizations should be collected simultaneously via the bunch-by-bunch switching of the electron polarization. Special attention must be paid to the luminosity and the average electron beam polarization monitoring. Precise enough luminosity monitoring can be provided with a dedicated low-angle Bhabha detector.

Data-driven monitoring of the average electron beam polarization provides the best control of the systematic uncertainty. We developed a method of the average beam polarization monitoring based on the angular analysis of the~$\jpsi\to [\Lambda\to p\pi^-][\lambar\to\pbar\pi^+]$ decay. The~$5D$ differential cross section for this decay in eq.~\eqref{eq:Wdistr} is derived taking into account the polarization of the electron beam. The expected statistical precision for the average polarization~$\pe$ does not limit the precision of the weak mixing angle measurement.

Longitudinal polarization of the electron beam significantly increases sensitivity to the~$\Lambda$ baryon form factors and $CP$ symmetry breaking in the~$\Lambda\to p\pi^-$ decay. There is no doubt that there are many applications of the beam polarization in studies of baryons. Cascade decays like~$\Xi^-\to[\Lambda\to p\pi^-]\pi^-$ and~$\Lambda_c^+\to[\Lambda\to p\pi^-]\pi^+$ are also sensitive to the beam polarization and should be considered in this context. 

\acknowledgments
The authors are grateful to I.A.~Koop for the discussions about technical aspects of making a polarized electron beam and to A.I.~Milstein for the discussions about calculation of differential distributions of the produced particles. Part of this work, the feasibility study (section~\ref{sec:feas}), [performed by V.V.] was supported by the Russian Science Foundation (grant 19-72-20114). Part of this work, appendix A, [performed by A.G.] was supported by the
Russian Fund of Basic Research (grant 19-02-00690).

%%%%%%%%%%%%%%%%%%%%
%%%%  APPENDIX  %%%%
%%%%%%%%%%%%%%%%%%%%
\newpage
\appendix
%================================================
% \section%[Reduced matrix element squared of e+e- -> Lambda (-> p pi+) anti-Lambda(-> anti-p pi-)]
% {\boldmath Reduced matrix element squared of \texorpdfstring{\label{app:matrix}
\section{Reduced matrix element squared}\label{app:matrix}
In this section we calculate the reduced matrix element squared for the $e^+e^-\to \jpsi \to[\Lambda\to p\pi^-][\lambar\to \pbar\pi^+]$ process depicted in figure \ref{fig:diagram}. We use kinematic variables defined in eq.~(\ref{kinematics}), which have the properties:
\begin{equation}\label{kinematicap}
l p \equiv l_{1}p_{1}=l_{2}p_{2}=(m_{\Lambda}^{2}+m_p^{2}-m_{\pi}^{2})/{2},\quad PQ=0,\quad k_{+}Q=-k_{-}Q.
\end{equation}
For brevity we keep both notations of the form factors $G^\psi_{M,E}$ and $G^\psi_{1,2}$ related as follows:
\begin{equation}
G_M^\psi =G^\psi_1,\quad
G_E^\psi=G^\psi_1-\frac{Q^2}{4\mlam^2}G^\psi_2,\quad
G^\psi_2=\frac{4\mlam^2}{Q^2}\left(G_M^\psi-G_E^\psi\right).
 \end{equation}
%We will use the kinematic variables from eq.~(\ref{kinematics}), which we rewrite here for convenience:
%\begin{equation}\label{kinematicap}
%P\equiv k_{+}+k_{-}=p_{1}+p_{2},\quad Q\equiv p_{1}-p_{2},\quad
%s\equiv P^{2}=4m_{\Lambda}^{2}-Q^{2}, \quad l p \equiv l_{1}p_{1}=l_{2}p_{2}.
%\end{equation}
%It is obvious that $PQ=0$, $k_{+}Q=-k_{-}Q$, and
%$l p=(m_{\Lambda}^{2}+m_p^{2}-m_{\pi}^{2})/{2}$,
%where $m_{\Lambda}$ is the Lambda hyperon mass, $m_p$ is the proton
%mass, and $m_{\pi}$ is the pion mass.
It is also convenient to introduce the notations
\begin{equation}
\begin{split}
\operatorname{Vol}\left(l_1,l_2,l_{3},l_{4}\right) & =\varepsilon_{\alpha
_1\alpha_2\alpha_{3}\alpha_{4}}\,l_1^{\alpha_1}l_2^{\alpha_2}l_{3}^{\alpha_{3}}l_{4}^{\alpha_{4}}, \\
\operatorname{Vol}\left(l_1,l_2,l_{3},\mu\right) & = \varepsilon_{\alpha_1\alpha_2\alpha_{3}\mu}\,l_1^{\alpha_1}l_2^{\alpha_2}l_{3}^{\alpha_{3}}, \\
\operatorname{Vol}\left(l_1,l_2,\mu,\nu\right) & =\varepsilon_{\alpha
_1\alpha_2\mu\nu}\,l_1^{\alpha_1}l_2^{\alpha_2}.
\end{split}
\end{equation}

\subsection{Leptonic and hadronic tensors}
\label{subsec:lephadtens}
% The~$e^+e^-\to \jpsi\to\Lambda\lambar$ cross section is proportional to the matrix element squared and the phase space factor
% \begin{equation}
% \dd\sigma = \frac{\overline{|\mathcal{M}|^2}}{4I} (2\pi)^4\delta^{(4)}(k_+ +k_- -p_1-p_2)\frac{\dd^3p_1}{(2\pi)^3 2\epsilon_1}\frac{\dd^3p_2}{(2\pi)^3 2\epsilon_2},
% \end{equation}
% where $I=\sqrt{(k_+ k_-)^2-k^2_+ k^2_-} \approx s/2$ is the M{\o}ller flux factor.

We defined the reduced matrix element squared and averaged over the positron's polarizations and summed over the proton and antiproton's polarizations in eq. (\ref{LH}). It is equal to the convolution of the leptonic~$L^{\mu\nu}$ and hadronic~$H_{\nu\mu}$ tensors.
% \begin{equation}
% \overline{|\mathcal{M}|^2} = \frac{(4\pi \alpha^{\xi}_\jpsi) (4\pi \alpha_{g})}{(s-m^2_\jpsi)^2+m^2_\jpsi \Gamma^2_\jpsi} L^{\mu\nu} H_{\nu\mu}.\label{M}
% \end{equation}
%\begin{equation}
% \overline{|\mathcal{M}|^2} \propto 
% \overline{|\mathcal{M}_\mathrm{red}|^2}=L^{\mu\nu} H_{\nu\mu}=a+b\xi.\label{M}
%\end{equation}
The leptonic tensor reads
\begin{equation}
\begin{split}
L^{\mu\nu} & =\left[\barr{v}(k_+)\gamma^\nu u(k_-)\right]^\dag
\barr{v}(k_+)\gamma^\mu u(k_-)=\frac{1}{4} \operatorname{Tr}\left[\gamma^\nu\hat{k}_+ \gamma^\mu\hat{k}_- (1-\xi\gamma^{5})\right]= \\
& = k_+^\mu k_-^\nu + k_-^\mu k_+^\nu - \frac{s}{2} g^{\mu\nu} - 
\xi i\varepsilon^{\mu\nu\alpha\beta}k_{-\alpha}k_{+\beta},
\end{split}
\end{equation}
where $\gamma^5=i\gamma^0\gamma^1\gamma^2\gamma^3$ and $\varepsilon^{0123}=1=-\varepsilon_{0123}$. 
Note that the sign of $\varepsilon^{\mu\nu\alpha\beta}$ here is different from ref. \cite{Faldt2015}.

Suppose the ultrarelativistic positron beam is (partially) polarized. Then a positron  has the longitudinal with respect to its momentum component of the double average spin in its rest frame $\zeta_{\shortparallel}$ and the perpendicular component $\vec{\zeta}_{\bot}$. Therefore one can present the leptonic tensor in the form
\begin{equation}
\begin{split}\label{Appleptonictensorgen}
L^{\mu\nu} & =\frac{1}{4} \operatorname{Tr}\left[\gamma^\nu\hat{k}_+(1+\zeta_{\shortparallel}\gamma^5-  \gamma^5 \vec{\gamma}_{\bot}\vec{\zeta}_{\bot} ) \gamma^\mu\hat{k}_- (1-\xi\gamma^{5})\right]= \\
& = (1-\xi \zeta_{\shortparallel})\left(k_+^\mu k_-^\nu + k_-^\mu k_+^\nu - \frac{s}{2} g^{\mu\nu}\right) - 
(\xi-\zeta_{\shortparallel}) i\varepsilon^{\mu\nu\alpha\beta}k_{-\alpha}k_{+\beta}.
\end{split}
\end{equation}
One can see that the leptonic tensor does not depend on $\vec{\zeta}_{\bot}$.

The hadronic tensor has the form
\begin{equation}\label{H}
 \begin{split}
  H_{\nu\mu} & =\operatorname{Tr}\left\{\left(\hat{p}_1+m_\Lambda \right) \left[R_\Lambda -S_\Lambda \gamma_{5}\left(lp+m_\Lambda \hat{l}_1\right) \right] \left({G_M^\psi}\gamma_\mu - \frac{2\mlam}{Q^2}\left({G_M^\psi} -{G_E^\psi}\right)Q_\mu\right)\times \right. \\
  & \quad\times\left. \left(\hat{p}_2-m_\Lambda \right) 
  \left[\barr{R}_\Lambda +\barr{S}_\Lambda \gamma_{5}\left(lp-m_\Lambda \hat{l}_2\right)\right] \left({G_M^{\psi\ast}}\gamma_\nu - \frac{2\mlam}{Q^2}\left({G_M^{\psi\ast}} -{G_E^{\psi\ast}}\right)Q_\nu\right)\right\}.
 \end{split}
\end{equation}
We present it as a sum of the symmetric ($\barr{H}_{\nu\mu}$) and antisymmetric ($\widetilde{H}_{\nu\mu}$) parts:
\begin{equation}
  H_{\nu\mu}=\barr{H}_{\nu\mu}+\widetilde{H}_{\nu\mu},\quad\barr{H}_{\nu\mu}=
  \frac{H_{\nu\mu}+H_{\mu\nu}}{2},\quad\widetilde{H}_{\nu\mu}=
  \frac{H_{\nu\mu}-H_{\mu\nu}}{2}.
\end{equation}
The symmetric component was calculated before in~ref.~\cite{Faldt2015} while the antisymmetric one is new. Both parts of the hadronic tensor from eq.~\eqref{H} have the form
\begin{equation}
 \begin{split}
  \barr{H}_{\nu\mu} & = \barr{R}_\Lambda R_\Lambda\barr{H}_{\nu\mu}^{RR}
  +\barr{R}_\Lambda S_\Lambda \barr{H}_{\nu\mu}^{RS}
  +\barr{S}_\Lambda R_\Lambda \barr{H}_{\nu\mu}^{SR} + 
  \barr{S}_\Lambda S_\Lambda \barr{H}_{\nu\mu}^{SS}, \\
  \widetilde{H}_{\nu\mu} & = \barr{R}_\Lambda R_\Lambda \widetilde{H}_{\nu\mu}^{RR}
  +\barr{R}_\Lambda S_\Lambda \widetilde{H}_{\nu\mu}^{RS}
  +\barr{S}_\Lambda R_\Lambda \widetilde{H}_{\nu\mu}^{SR}
  +\barr{S}_\Lambda S_\Lambda \widetilde{H}_{\nu\mu}^{SS}.
 \end{split}
\end{equation}
Here
\begin{equation}
 \begin{split}
  \barr{H}_{\nu\mu}^{RR} & = 2\left\vert G^\psi_1\right\vert^2 
  \left(P_\mu P_\nu - P^2 g_{\mu\nu} - Q_\mu Q_\nu \right) + 2Q_\mu Q_\nu
  \left[2\operatorname{Re}\left(G^\psi_1 G^{\psi\ast}_2 \right) -
  \frac{Q^2 \left\vert G^\psi_2\right\vert^2}{4m_\Lambda^2}\right],\\
  \barr{H}_{\nu\mu}^{RS} & = -2\operatorname{Im}
  \left(G^\psi_1 G^{\psi\ast}_2\right)
  \Bigl[Q_\nu\operatorname{Vol}\left(l_1,p_1,p_2,\mu\right)
  +Q_\mu\operatorname{Vol}\left(l_1,p_1,p_2,\nu\right)\Bigr],\\
  \barr{H}_{\nu\mu}^{SR} & = -2\operatorname{Im}
  \left(G^\psi_1 G^{\psi\ast}_2\right)
  \Bigl[Q_\nu\operatorname{Vol}\left(l_2,p_1,p_2,\mu\right)
  +Q_\mu\operatorname{Vol}\left(l_2,p_1,p_2,\nu\right)\Bigr],\\
  \barr{H}_{\nu\mu}^{SS} & = -\left\vert G^\psi_2\right\vert^2 Q_\mu Q_\nu
  \left(\frac{Q^2}{2m_\Lambda^2}
  \left[(lp)^2-m_\Lambda^2\left(l_1l_2\right)\right]
  +\left(l_1 Q\right) \left(l_2 Q\right) \right)+ \\
  & \quad +\left\vert G^\psi_1\right\vert^2
  \biggl\{-2\left(m_\Lambda^2\left(l_1l_2\right) + (lp)^2\right) \left(P_\mu P_\nu - sg_{\mu\nu} - Q_\mu Q_\nu \right)- \\
  & \quad -4m_\Lambda^2 \Bigl[\left(l_1 P\right) \left(l_2 P\right) g_{\mu\nu}+
  \frac{s}{2}\left(l_{2\mu}l_{1\nu}+l_{1\mu}l_{2\nu}\right)- \\
  & \quad -\left(l_1 P\right) (l_{2\nu}p_{1\mu}+l_{2\mu}p_{1\nu})-\left(l_2 P\right) (l_{1\nu}p_{2\mu}+l_{1\mu}p_{2\nu})\Bigr]\biggr\}+ \\
  & \quad +\operatorname{Re}\left(G^\psi_1 G^{\psi\ast}_2\right) 
  \Bigl[\left(l_{1\nu}Q_\mu+l_{1\mu}Q_\nu\right) \left(2m_\Lambda^2 (Pl_2)-(pl)s\right)- \\
  & \quad -\left(l_{2\nu}Q_\mu+l_{2\mu}Q_\nu\right) \left(2m_\Lambda^2(Pl_1)-(pl)s\right) -4m_\Lambda^2 Q_\mu Q_\nu\left(l_1l_2\right)+ \\
  & \quad +2\left(p_{1\nu}Q_\mu+p_{1\mu}Q_\nu\right)
  \left(l_1p_2\right)-2\left(p_{2\nu}Q_\mu+p_{2\mu}Q_\nu\right)
  \left(l_2p_1\right)\Bigr].
 \end{split}
\end{equation}
This expression coincides with the symmetric part of the hadronic tensor from ref.~\cite{Faldt2015}.
\begin{equation}
 \begin{split}
  \widetilde{H}_{\nu\mu}^{RR} & = 0,\\ 
  \widetilde{H}_{\nu\mu}^{RS} & = 4i\left\vert G^\psi_1\right\vert^2
  \Bigl[(lp)\operatorname{Vol}\left(p_1,p_2,\mu,\nu\right) - m_\Lambda^2 \operatorname{Vol}\left(l_1,P,\mu,\nu\right)\Bigr]+ \\
  & \quad +2i\operatorname{Re}\left(G^\psi_1 G^{\psi\ast}_2\right)
  \Bigl[Q_\nu \operatorname{Vol}\left(l_1,p_1,p_2,\mu\right) - Q_\mu \operatorname{Vol}\left(l_1,p_1,p_2,\nu\right) \Bigr], \\
  \widetilde{H}_{\nu\mu}^{SR} & = -4i\left\vert G^\psi_1\right\vert^2
  \Bigl[(lp)\operatorname{Vol}\left(p_1,p_2,\mu,\nu\right) +m_\Lambda^2\operatorname{Vol}\left(l_2,P,\mu,\nu\right) \Bigr]+ \\
  & \quad +2i\operatorname{Re}\left(G^\psi_1G^{\psi\ast}_2\right)
  \Bigl[Q_\nu\operatorname{Vol}\left(l_2,p_1,p_2,\mu\right) -Q_\mu\operatorname{Vol}\left(l_2,p_1,p_2,\nu\right) \Bigr], \\
  \widetilde{H}_{\nu\mu}^{SS} & = i\operatorname{Im}\left(G^\psi_1 G^{\psi\ast}_2\right) 
  \biggl\{-(lp)\left(l_1p_2-l_2p_1\right)\left(P_\nu Q_\mu-P_\mu Q_\nu\right)- \\
  & \quad -\left(l_{2\nu}Q_\mu-l_{2\mu}Q_\nu\right) 
  \Bigl[\left(2m_\Lambda^2-Q^2\right)(lp)-2m_\Lambda^2\left(l_1p_2\right) \Bigr]+ \\
  & \quad +\left(l_{1\nu}Q_\mu-l_{1\mu}Q_\nu\right) 
  \Bigl[\left(2m_\Lambda^2-Q^2\right)(lp)-2m_\Lambda^2\left(l_2p_1\right)\Bigr]\biggr\}.
 \end{split}
\end{equation}

\subsection{Two parts of the reduced matrix element squared}\label{app:matrix2}
The reduced matrix element squared defined in eq.~\eqref{LH} has the unpolarized part denoted $a$ and the part named $b$ responsible for the polarization. We present the quantities $a$ and $b$ in the following form
\begin{equation}
 \begin{split}
   a & = \barr{R}_\Lambda R_\Lambda a^{RR} + \barr{R}_\Lambda S_\Lambda a^{RS} + \barr{S}_\Lambda R_\Lambda a^{SR} + \barr{S}_\Lambda S_\Lambda a^{SS},\\
   b & = \barr{R}_\Lambda R_\Lambda b^{RR} + \barr{R}_\Lambda S_\Lambda b^{RS} + \barr{S}_\Lambda R_\Lambda b^{SR} + \barr{S}_\Lambda S_\Lambda b^{SS}.
 \end{split}
\end{equation}
Here
\begin{equation}
 \begin{split}
  a^{RR} & =\left\vert G^\psi_1\right\vert^2
  \left(4\left(k_+ Q\right)^2 + s\left(4m_\Lambda^2+s\right) \right)- \\
  & \quad -\left(2\left(k_+ Q\right)^2+\frac{s}{2}Q^2\right)
  \left[4\operatorname{Re}\left(G^\psi_1G^{\psi\ast}_2\right) -\frac{Q^2}{2m_\Lambda^2}\left\vert G^\psi_2\right\vert^2\right],\\
  a^{RS} & = 4\operatorname{Im}\left(G^\psi_1 G^{\psi\ast}_2\right) 
  \left(k_+ Q\right)\operatorname{Vol}\left(k_- - k_+, l_1, p_1, p_2\right),\\
  a^{SR} & = 4\operatorname{Im}\left(G^\psi_1 G^{\psi\ast}_2\right)
  \left(k_+ Q\right)\operatorname{Vol}\left(k_- - k_+, l_2, p_1, p_2\right),\\
  a^{SS} & =\left\vert G^\psi_2\right\vert^2
  \left(2\left(k_+ Q\right)^2+\frac{s}{2}Q^2\right) \left[\frac{Q^2}{2m_\Lambda^2}(lp)^2+\left(l_1 Q\right) \left(l_2 Q\right) -\frac{Q^2}{2}\left(l_1 l_2\right)\right]+ \\
  & \quad +2\left\vert G^\psi_1\right\vert^2\biggl\{m_\Lambda^2 s
  \left(2\left(k_- l_1\right)\left(k_- l_2\right) + 2\left(k_+ l_1\right)
  \left(k_+ l_2\right) - 2\left(l_1 p_2\right) \left(l_2 p_1\right) -\frac{1}{2}Q^2\left(l_1 l_2\right)\right)- \\
  & \quad -2m_\Lambda^2\left(k_+ Q\right) 
  \Bigl[\left(l_1 l_2\right)\left(k_+ Q\right) - 2\left(k_- l_2\right)\left(k_+ l_1\right) + 2\left(k_- l_1\right) \left(k_+ l_2\right) \Bigr]- \\
  & \quad -\frac{1}{2}(lp)^2\left(4\left(k_+ Q\right)^2 + P^4\right)\biggr\}+ \\
  & \quad +\operatorname{Re}\left(G^\psi_1G^{\psi\ast}_2\right) 
  \biggl\{\left(k_+ Q\right)^2\Bigl[8m_\Lambda^2\left(l_1 l_2\right) -4(lp)\left(l_1 p_2 + l_2 p_1\right) \Bigr]+ \\
  & \quad +\left(k_+ Q\right)\biggl(2\left(2m_\Lambda^2-s\right) (lp)
  \left(k_+ - k_-,l_2-l_1\right)+ \\ 
  & \quad +4m_\Lambda^2\Bigl[\left(l_1p_2\right)
  \left(k_+ - k_-,l_2\right) -\left(l_2p_1\right) 
  \left(k_+ - k_-,l_1\right) \Bigr] \biggr)+ \\
  & \quad + 2s\left[\left(s-2m_\Lambda^2\right)(lp)^2 + m_\Lambda^2\Bigl[2\left(l_1p_2\right) \left(l_2p_1\right)
  +Q^2\left(l_1l_2\right)\Bigr] - 2m_\Lambda^2(lp)
  \left(l_1p_2 + l_2p_1\right) \right] \biggr\}.
 \end{split}
\end{equation}
Here $a=-A-sB/2$ with $A$ and $B$ from eqs.~(40)\,--\,(47) in ref.~\cite{Faldt2015}.
\begin{equation}
 \begin{split}
  b^{RR} & = 0,\\
  b^{RS} & = 4\left\vert G^\psi_1\right\vert^2
  \Bigl((lp)\left(k_+ Q\right) - m_\Lambda^2 \left(k_+ - k_-, l_1\right)\Bigr) s + \\
  &\quad +\operatorname{Re}\left(G^\psi_1G^{\psi\ast}_2\right) 
  \Bigl[2\left(k_+ Q\right) \Bigl(l_1P-2(lp)\Bigr) + Q^2\left(k_+ - k_-, l_1\right) \Bigr] s,\\
  b^{SR} & = -4\left\vert G^\psi_1\right\vert^2
  \Bigl((lp)\left(k_+ Q\right) + m_\Lambda^2 \left(k_+ - k_-, l_2\right) \Bigr) s + \\
  &\quad +\operatorname{Re}\left(G^\psi_1G^{\psi\ast}_2\right)
  \Bigl[2\left(k_+ Q\right) \Bigl(2(lp)-l_2P\Bigr) + Q^2\left(k_+ - k_-, l_2\right) \Bigr] s, \\
  b^{SS} & = 2\operatorname{Im}\left(G^\psi_1G^{\psi\ast}_2\right)
  \Bigl[\Bigl(2m_\Lambda^2\left(l_2P\right) - s(lp)\Bigr) 
  \operatorname{Vol}\left(k_-, k_+, l_1, Q\right)+ \\
  &\quad +\Bigl(s(lp)-2m_\Lambda^2\left(l_1P\right) \Bigr)
  \operatorname{Vol}\left(k_-, k_+, l_2, Q\right) \Bigr].
 \end{split}
\end{equation}

%============================
\section{Distributions in \cms frame}
\label{app:cmsdistrib}
The distribution in eq.~\eqref{eq:3d-csec} is expressed in terms of two frames: the \cms frame (angle $\theta$) and the $\Lambda$ frame ($d\Omega_1=\sin\theta_1 d\theta_1 d\phi_1$). To express the angular distributions only through the \cms frame variables, we introduce $\theta_1^{(0)}$ and $\phi_1^{(0)}$ as the polar and azimuth angles of the proton in the \cms frame ($d \Omega^{(0)}_1 =\sin \theta_1^{(0)} d \theta_1^{(0)} d \phi_1^{(0)}$), whereas $\theta$ and $\phi$ are the polar and azimuth angles of the $\Lambda$-hyperon in this frame ($d\Omega = \sin\theta\, d\theta\, d\phi$). 
\subsection{Two-valuedness in \cms frame}
\label{app:twovaluedness}

Let us consider the transformation of the proton's % (resulting from the $\Lambda$-hyperon decay) 
momentum from the $\Lambda$ frame to the \cms frame. The $\Lambda$ frame is defined according to eqs. (6.49) -- (6.51) from ref.~\cite{Faldt2017}. The proton's momentum $l^{(\Lambda)}_p$ and energy $\epsilon_p^{(\Lambda)}$ in this frame stay constant:
\begin{equation}
  l^{(\Lambda)}_p=\frac{1}{2m_\Lambda}\sqrt{(m_\Lambda - m_p - m_\pi)
  (m_\Lambda + m_p - m_\pi)(m_\Lambda - m_p + m_\pi)
  (m_\Lambda + m_p + m_\pi)},
\end{equation}
and
\begin{equation}
  \epsilon_p^{(\Lambda)}=\sqrt{l^{(\Lambda)2}_p + m^2_p} = (m_\Lambda^2 + m^2_p - m^2_\pi)/(2m_\Lambda).
\end{equation}
Using experimental values from ref.~\cite{pdg}: $m_\Lambda=1115.7~\mev$,
$m_p=938.2721~\mev$, and $m_\pi=139.57~\mev$, we obtain that the proton is quite nonrelativistic in the~$\Lambda$ frame:
\begin{equation}
  \beta_p^{(\Lambda)}=\frac{l_p^{(\Lambda)}}{\epsilon_p^{(\Lambda)}}\approx 0.11.
\end{equation}
In the \cms frame ($x_0$, $y_0$, $z_0$) the~$z_0$ axis is directed along the momentum of the electron (i.e. along~$\bf{k}_-$) and the~$e^+e^- \to \Lambda\lambar$ scattering plane is inclined at the angle~$\phi$ to the~$x_0 z_0$ coordinate plane, as is shown in figure~\ref{fig:axesChoice}.

Thus we have the following transformation including the Lorentz boost along the~$z$-axis, the rotation of the coordinate system about the~$y$-axis at the angle~$\theta$, and the rotation of the coordinate system about the new~$z$-axis ($z_0$-axis) at the angle~$\pi-\phi$:

\begin{equation}\label{transfomLambdato0}
\left(
\begin{array}{c}
 \epsilon_p^{(0)} \\
 l_p^{(0)} \sin \text{$\theta_1^{(0)} $} \cos \text{$\phi_1^{(0)} $} \\
 l_p^{(0)} \sin \text{$\theta_1^{(0)} $} \sin \text{$\phi_1^{(0)} $} \\
 l_p^{(0)} \cos \text{$\theta_1^{(0)} $}
\end{array}
\right)=\hat R \left(
\begin{array}{c}
 \epsilon_p^{(\Lambda)} \\
l^{(\Lambda)}_p \sin \text{$\theta_1 $} \cos \text{$\phi_1 $} \\
l^{(\Lambda)}_p \sin \text{$\theta_1 $} \sin \text{$\phi_1 $} \\
l^{(\Lambda)}_p \cos \text{$\theta_1 $}
\end{array}
\right),
\end{equation}
where $\epsilon_p^{(0)}$ and $ l_p^{(0)}$ are the proton's energy and momentum in the \cms frame. The transformation $\hat R$ reads
\begin{equation}
\hat R= \hat R_{z(\phi)} \hat R_{y(\theta)}\hat
R_{0(\beta_\Lambda)}= \left(
\begin{array}{cccc}
 \gamma_\Lambda & 0 & 0 & \beta_\Lambda \gamma_\Lambda \\
 \beta_\Lambda \gamma_\Lambda \sin\theta \cos\phi & -\cos\theta \cos\phi
 & \sin\phi & \gamma_\Lambda \sin\theta \cos\phi \\
 \beta_\Lambda \gamma_\Lambda \sin\theta \sin\phi & -\cos\theta \sin\phi
 & -\cos\phi & \gamma_\Lambda \sin\theta \sin\phi \\
 \beta_\Lambda \gamma_\Lambda \cos\theta & \sin\theta & 0 & \gamma_\Lambda \cos\theta
\end{array}
\right),
\end{equation}
where $\beta_\Lambda=\sqrt{1-4m_\Lambda^2/s}$ and
$\gamma_\Lambda=\sqrt{s}/(2m_\Lambda)$ are the boost parameters ($\beta_\Lambda\approx 0.69$ and $\gamma_\Lambda \approx 1.39$ for $s=m_\jpsi^2$), and
\begin{equation}\label{matrixes}
\begin{split}
\hat R_{0(\beta_\Lambda)} & =\left(
\begin{array}{cccc}
 \gamma_\Lambda & 0 & 0 & \beta_\Lambda \gamma_\Lambda \\
 0 & 1 & 0 & 0 \\
 0 & 0 & 1 & 0 \\
 \beta_\Lambda \gamma_\Lambda & 0 & 0 & \gamma_\Lambda
\end{array}
\right), \\
\hat R_{y(\theta)} & =\left(
\begin{array}{cccc}
 1 & 0 & 0 & 0 \\
 0 & \cos \theta & 0 & -\sin \theta \\
 0 & 0 & 1 & 0 \\
 0 & \sin \theta & 0 & \cos \theta
\end{array}
\right), \\ 
\hat R_{z(\phi)} & =\left(
\begin{array}{cccc}
 1 & 0 & 0 & 0 \\
 0 & -\cos \phi & \sin \phi & 0 \\
 0 & -\sin \phi & -\cos \phi & 0 \\
 0 & 0 & 0 & 1
\end{array}
\right).
\end{split}
\end{equation}
Eq.~\eqref{transfomLambdato0} allows one to find the energy
$\epsilon_p^{(0)}$ and the momentum $l_p^{(0)}$ as functions of the
angles $\theta$, $\phi$ and $\theta_1^{(0)}$, $\phi_1^{(0)}$.
For $s=m_\jpsi^2$ we have $\beta_\Lambda>\beta_p^{(\Lambda)}$, i.e. $s>(m_\Lambda^2+m_p^2-m^2_\pi)^2/m_p^2$. Therefore for fixed $\theta_1^{(0)}$ and $\phi_1^{(0)}$ there are two solutions for $\epsilon_p^{(0)}$ and $l_p^{(0)}$:
\begin{equation}\label{energyp0mt}
\epsilon_p^{(0)}=\frac{\epsilon_p^{(\Lambda)}\pm g \beta_\Lambda 
\sqrt{\epsilon_p^{(\Lambda)\,2}-\gamma_\Lambda^2
m_p^2(1-\beta_\Lambda^2 g^2)}}{\gamma_\Lambda (1-\beta_\Lambda^2 g^2)}, \quad
l_p^{(0)}=\frac{g \beta_\Lambda \epsilon_p^{(\Lambda)}\pm
\sqrt{\epsilon_p^{(\Lambda)\,2}-\gamma_\Lambda^2
m_p^2(1-\beta_\Lambda^2 g^2)}}{\gamma_\Lambda (1-\beta_\Lambda^2 g^2)},
\end{equation}
where we used the following function $g$:
\begin{equation}\label{g-functionmt}
g=g(\theta, \phi;\theta_1^{(0)},\phi_1^{(0)})= \cos \theta \cos
\theta_1^{(0)}+ \sin \theta \sin \theta_1^{(0)} \cos
(\phi-\phi_1^{(0)}).
\end{equation}
It is obvious that $g={\bf n}_\Lambda \cdot {\bf n}^{(0)}_p$, where ${\bf n}_\Lambda$ and ${\bf n}^{(0)}_p={\bf l}_p^{(0)}/l_p^{(0)}$ are the unit vectors along the $\Lambda$-hyperon's and the proton's momenta in the \cms frame, respectively. Moreover, the acceptable angles
$\theta_1^{(0)}$ and $\phi_1^{(0)}$ in the \cms frame obey the inequality, which ensures that the expression under the root in eq.~\eqref{energyp0mt} is not negative: \begin{equation}\label{regionmt}
\left. g(\theta,\phi; \theta_1^{(0)},\phi_1^{(0)}) \geq
g_{thr}(s)=\sqrt{1-\frac{(m_\Lambda^2+m_p^2-m^2_\pi)^2}{m^2_p s}}\middle/\sqrt{1-\frac{4m_\Lambda^2}{s}}=\sqrt{1-\delta^2(s)}. \right.
\end{equation}
Here we introduce the parameter $\delta(s)$ defined in eq.~(\ref{eq:delta1}).
%\begin{equation}
%\delta(s)=\frac{l^{(\Lambda)}_p}{m_p \beta_\Lambda \gamma_\Lambda}=\frac{2 m_\Lambda}{\sqrt{s-4m_\Lambda^2}} %\frac{l^{(\Lambda)}_p}{m_p}.
%\end{equation}
To express the squared amplitude, e.g. eq.~\eqref{eq:Wdistr}, in terms of the angles $\theta$, $\phi$,
$\theta_1^{(0)}$, and $\phi_1^{(0)}$ in the \cms frame, we use the following relations:
\begin{equation}\label{coshteta1}
\cos \theta_1 =\gamma_\Lambda \frac{g l_p^{(0)}- \beta_\Lambda 
\epsilon_p^{(0)}}{l^{(\Lambda)}_p}=\frac{(g^2-1) \beta_\Lambda 
\gamma_\Lambda^2 \epsilon_p^{(\Lambda)}\pm g
\sqrt{\epsilon_p^{(\Lambda)\,2}-\gamma_\Lambda^2
m_p^2(1-\beta_\Lambda^2 g^2)}}{l^{(\Lambda)}_p
\gamma_\Lambda^2(1-\beta_\Lambda^2 g^2)},
\end{equation}
\begin{equation}\label{sinphi1}
\sin \phi_1 =\frac{l_p^{(0)} \sin \theta_1^{(0)}\sin
(\phi-\phi_1^{(0)})}{l^{(\Lambda)}_p \sin \theta_1}, \qquad \cos
\phi_1= \frac{l_p^{(0)} g'}{l^{(\Lambda)}_p\sin \theta_1},
\end{equation}
where $g$ is defined in eq.~\eqref{g-functionmt}, $\sin
\theta_1=+\sqrt{1-\cos^2 \theta_1}$, and the function $g'$ is defined as follows:
\begin{equation}\label{gprime-functionmt}
g'=g'(\theta, \phi;\theta_1^{(0)},\phi_1^{(0)})= \sin \theta \cos
\theta_1^{(0)}-\cos\theta \sin \theta_1^{(0)} \cos
(\phi-\phi_1^{(0)}).
\end{equation}

As a result, the angle $\phi_1$ can be restored from $\theta$, $\phi$,
$\theta_1^{(0)}$, and $\phi_1^{(0)}$ without ambiguity.

% \subsection[Differential cross section of the process e+e- -> Lambda anti-Lambda -> anti-Lambda p pi- in \cms frame] {Differential cross section of the process $e^+e^- \to \gamma^* (Z^*) \to \jpsi \to \Lambda \bar
% \Lambda \to \bar \Lambda p \pi^- $ in \cms frame}
\subsection{Differential cross section in \cms frame}\label{app:Difcs}
Let us now consider the special case when the $\lambar$-hyperon
is not detected, i.e. the process $e^+e^- \to \gamma^* (Z^*) \to \jpsi \to \Lambda \bar
\Lambda \to \bar \Lambda p \pi^- $ with the polarized electron
beam and the unpolarized positron one.
Its differential cross section has the form
\begin{equation}
d \sigma =\dfrac{\overline{\left|{\cal M}\right|^2}}{4 I} d\rho_f.
\end{equation}
Here the flux factor $I=\sqrt{(k_+ k_-)^2 - k_+^2 k_-^2}\approx
s/2$, and the Lorentz invariant phase space element reads
\begin{equation}
\begin{split}
d\rho_f & =(2\pi)^4 \delta^{(4)}(k_+ + k_- - p_2 - l_1 - q_1)\frac{d^3 p_2}{(2\pi)^3
2\epsilon_2} \frac{d^3 l_1}{(2\pi)^3 2 \epsilon^{(0)}_p} \frac{d^3 q_1}{(2\pi)^3 2\epsilon^{(0)}_\pi}=\\
& = \int \frac{d s_1}{(2\pi)} \int \frac{d^3 p_1}{(2\pi)^3 
 2\epsilon_1} (2\pi)^4 \delta^{(4)}(k_+ +k_- -p_1-p_2) \frac{d^3 p_2}{(2\pi)^3
2\epsilon_2} \times \\
& \quad \times (2\pi)^4 \delta^{(4)}(p_1-l_1-q_1)\frac{d^3 l_1}{(2\pi)^3 
2\epsilon^{(0)}_p} \frac{d^3 q_1}{(2\pi)^3 2\epsilon^{(0)}_\pi},
\end{split}
\end{equation}
where $s_1=p^2_1$ and the pion's energy in the \cms frame $\epsilon^{(0)}_\pi=\sqrt{s}/2 -
\epsilon^{(0)}_p$.

In section \ref{subsec:helampl} we denoted the helicities of $\Lambda$, $\lambar$, proton and antiproton by $\lambda_1$, $\lambda_2$, $\lambda'_1$, and $\lambda'_2$ correspondingly keeping $\xi$ for the double helicity of the initial electron.
The squared amplitude summed over polarizations of the final particles 
$\bar\Lambda (p_2, \lambda_2)$ and $p(l_1,\lambda'_1)$ can be rewritten as
\begin{equation}\label{amplitude}
 \begin{split}
  \overline{\left|{\cal M}\right|^2}&=\frac{(4\pi \alpha^{\xi}_\jpsi) (4\pi \alpha_{g})}{(s-m^2_\jpsi)^2+m^2_\jpsi \Gamma^2_\jpsi} 
  \frac{\pi\delta(s_1-m_\Lambda^2)}{m_\Lambda \Gamma_\Lambda}\times\\
  &\quad\times\frac{1}{2}\sum_{\lambda_2,\lambda'_1}\Big|j_{(e)\mu}\sum_{\lambda_1}
  {\cal M}^\mu_{\Lambda \lambar}(\lambda_1,\lambda_2)
  {\cal M}_{(\Lambda)}(\lambda_1,\lambda'_1)\Big|^2,
 \end{split}
\end{equation}
where the factors $\alpha_{g}$ and $\alpha^{\xi}_\jpsi$ are defined in eqs.~\eqref{alphas} and \eqref{eq:alphaxi}, respectively, and $\Gamma_\Lambda$ is the total width of the $\Lambda$-hyperon. In eq.~\eqref{amplitude} we use the propagator on-shell approximation for the $\Lambda$-hyperon
\begin{equation}
  \frac{1}{(p^2_1-m_\Lambda^2)^2+m_\Lambda^2 \Gamma^2_\Lambda}\approx \frac{\pi}{m_\Lambda \Gamma_\Lambda} \delta(p^2_1-m_\Lambda^2), \qquad \Gamma_\Lambda \ll m_\Lambda.
\end{equation}
The factor $\frac{1}{2}$ in eq.~\eqref{amplitude} arises because the positron is unpolarized whereas the electron has the given double helicity $\xi=\pm 1$.
%\begin{equation}
%\left(\frac{4\pi \alpha_{em}}{s}\right)^2 \to \frac{(4\pi %\alpha^{\xi}_\jpsi) (4\pi \alpha_{g})}{(s-m^2_\jpsi)^2+m^2_\jpsi %\Gamma^2_\jpsi},
%\end{equation}
%where $\alpha_\jpsi=e^2_\jpsi/(4\pi)$ is the constant of the
%decay $\jpsi\to \gamma^* \to e^+e^-$, and
%$\alpha_{g}=e^2_{g}/(4\pi)$ is the constant of the decay
%$\jpsi\to \Lambda \lambar$, i.e.,
%\begin{equation}\label{alphas}
%&&\alpha_\jpsi=3 \Gamma_{\jpsi \to e^+e^-}/m_\jpsi, \nonumber \\&&
%\alpha_{g}=3
%\left((1+2m_\Lambda^2/m^2_\jpsi)\sqrt{1-4m_\Lambda^2/m^2_\jpsi}\right)^{-1}
%\Gamma_{\jpsi \to \Lambda \lambar}/m_\jpsi.
%\end{equation}
eq.~\eqref{amplitude} can also be obtained (up to the overall factor) from the
distribution in eq.~\eqref{eq:Wdistr} integrating the latter over $ \theta_2$
and $\phi_2$. Using eqs.~\eqref{jmue}, \eqref{Mlambda}, and
\eqref{Mpip}, we get
\begin{equation}
\begin{split}
\sum_{\lambda_2,\lambda'_1}\Big|j_{(e)\mu} & \sum_{\lambda_1}
{\cal M}^\mu_{\Lambda \lambar}(\lambda_1,\lambda_2)
{\cal M}_{(\Lambda)}(\lambda_1,\lambda'_1)\Big|^2 = 2 s R_\Lambda \left(s\left\vert G^\psi_M\right\vert^2+4m_\Lambda^2\left\vert
G^\psi_E\right\vert^2\right)\times \\ 
& \times \Bigg\{1+\alpha
\cos^2 \theta + \alpha_1 \sqral \sindphi \sin\theta \cos\theta \sin\theta_1 \sin\phi_1 + \\ 
& \quad + \xi \alpha_1 \Big[
\sqral \cosdphi \sin\theta \sin\theta_1 \cos\phi_1 + (1+\alpha) 
\cos\theta \cos\theta_1 \Big]
\Bigg\}.
\end{split}
\end{equation}
Since this quantity is Lorentz invariant, one can rewrite it in the \cms frame using eqs.~\eqref{coshteta1} and \eqref{sinphi1} to obtain the differential cross section of the $e^+e^- \to \gamma^* (Z^*)
\to \jpsi \to \Lambda \bar \Lambda \to \bar \Lambda p \pi^- $ process
in the \cms frame:
\begin{equation}\label{sigma1mt}
\begin{split}
\frac{d\sigma}{d \Omega^{(0)}_1 d \Omega} & = \mathcal{B}(\Lambda \to p
 \pi^-)\, \frac{s\,m_\Lambda \beta_\Lambda \,\alpha^{\xi}_\jpsi\alpha_{g} H\bigl(g-g_{thr}(s)\bigr)}{4\pi l^{(\Lambda)}_p\Big[(s-m^2_\jpsi)^2+m^2_\jpsi \Gamma^2_\jpsi\Big]} 
 \frac{\left|G^\psi_M\right|^2}{1+\alpha}
 \sum_{\pm} \frac{l_p^{(0)}}{\sqrt{s}/2-\epsilon_p^{(0)}}\times \\ 
 &\times \left\{1 + \alpha\cos^2\theta
 + \alpha_1 \sqral \sindphi \sin\theta \cos\theta \frac{l_p^{(0)}}{l^{(\Lambda)}_p}\sin
 \theta_1^{(0)}\sin (\phi-\phi_1^{(0)})+ \right. \\
 & \left. \quad  +\xi \alpha_1 
 \left[\sqral \cosdphi \sin\theta\, \frac{g'\, l_p^{(0)}}{l^{(\Lambda)}_p} + (1+\alpha)\cos\theta\,\gamma_\Lambda \frac{g\, l_p^{(0)} - \beta_\Lambda \epsilon_p^{(0)}}{l^{(\Lambda)}_p}\right] \right\}.
\end{split}
\end{equation} % $B=4\pi {l^{(\Lambda)}_p}/{m_\Lambda}\approx 1.13$
Here we use the notations from eqs.~\eqref{g-functionmt} and \eqref{gprime-functionmt}, and
\begin{equation}
    \mathcal{B}(\Lambda \to p \pi^-)=\Gamma_{\Lambda \to p
\pi^-}/\Gamma_\Lambda = (63.9 \pm 0.5) \%,\quad \Gamma_{\Lambda \to p
\pi^-}= R_\Lambda l^{(\Lambda)}_p/(8 \pi m_\Lambda^2).
\end{equation}
In eq.~\eqref{sigma1mt} the sum $\sum_{\pm}$ goes over two branches of
the solutions from eq.~\eqref{energyp0mt}. 
The Heaviside step function $H$ in the first line of eq.~\eqref{sigma1mt} means that we consider only the acceptable
proton's angles given by eq.~\eqref{regionmt}. 

%For the large invariant $s=(k_+ +k_-)^2$ the proton's angles $\theta_1^{(0)}$ and $\phi_1^{(0)}$ in the \cms frame are very close to the $\Lambda$-hyperon's angles $\theta$ and $\phi$ (see eq.~\eqref{regionsimpl2} in the Appendix, section~\eqref{app:protdistrib}) and it is easy to integrate eq.~\eqref{sigma1mt} over $\theta$ and $\phi$ using expansion in $\delta(s) \ll 1$. On the contrary, if $s$ is not large integration over $\theta$ and $\phi$ results in elliptic functions and is difficult for analysis. However, for $s=m^2_\jpsi$ one has sufficiently small $\delta(s_\jpsi) \thickapprox 0.111$.

%The expression \eqref{sigma1} coincides with one in the worknote
%\cite{Milstein} up to the sum notation $\sum_{\pm}$ and coefficient
%$1/2$.

\subsection{Angular distribution of protons in \cms frame at large energies}
\label{app:protdistrib}
For the large invariant $s$, when $\delta(s) \ll 1$, one can simplify eq.~\eqref{energyp0mt} considerably. 
In this case, as follows from eq.~\eqref{regionmt}, 
\begin{equation}
g={\bf n}_\Lambda \cdot {\bf n}^{(0)}_p \geq \sqrt{1-\delta^2(s)}\approx 1-\delta^2(s)/2,
\end{equation} i.e. the region for the acceptable $\Lambda$-hyperon's angles $\theta$ and $\phi$ reduces to a small neighborhood near the proton's angles $\theta^{(0)}_1$ and $\phi^{(0)}_1$. Introducing 
\begin{equation}
\delta\theta_1^{(0)}=\theta_1^{(0)}-\theta , \quad |\delta\theta_1^{(0)}|\ll 1,\qquad \delta \phi_1^{(0)}= \phi_1^{(0)}-\phi,\quad|\delta \phi_1^{(0)}|\ll 1,
\end{equation}
we get for the acceptable region
\begin{equation}\label{regionsimpl2}
\delta\theta_1^{(0)\,2}+ \sin\theta^{(0)}_1 
\sin\left(\theta^{(0)}_1-\delta\theta_1^{(0)}\right) \delta\phi_1^{(0)\,2} \leq \delta^2(s) \ll 1,
\end{equation}
which is close to the interior of an ellipse when the 
proton's polar angle is not small, i.e. $\theta^{(0)}_1 \gg \delta(s)$. Numerically, our approximation  works well down to $\theta^{(0)}_1 \gtrsim 2 \delta(s)$, which we assume hereafter.

For $\delta(s) \ll 1$ we can integrate eq.~\eqref{sigma1mt} over the $\Lambda$-hyperon's angles $\theta$ and $\phi$ or,
equivalently, $\delta\theta_1^{(0)}$ and $\delta \phi_1^{(0)}$ using expansion in $\delta(s)$. The latter variables range over the
region defined in eq.~\eqref{regionsimpl2} with the accuracy of ${\cal O}(\delta^4(s))$.
During the integration the proton's angles $\theta_1^{(0)}$ and
$\phi_1^{(0)}$ are kept fixed. 

If $s$ is not large the integration over $\theta$ and $\phi$ results in 
elliptic functions and is difficult for analysis. However, even for
the $\jpsi$ energy ($\sqrt{s}=3096.9$ MeV) $\delta(s)$ is small enough: $\delta(s)\approx
0.111$. As follows from eq.~(\ref{eq:delta1}), $\delta(s)$ is proportional to the numerically small coefficient $l^{(\Lambda)}_p/m_p \approx 0.107$, which shows that the proton is nonrelativistic in the $\Lambda$ frame.

\begin{figure}[ptb]%
\centering
\includegraphics[width=0.7\textwidth]{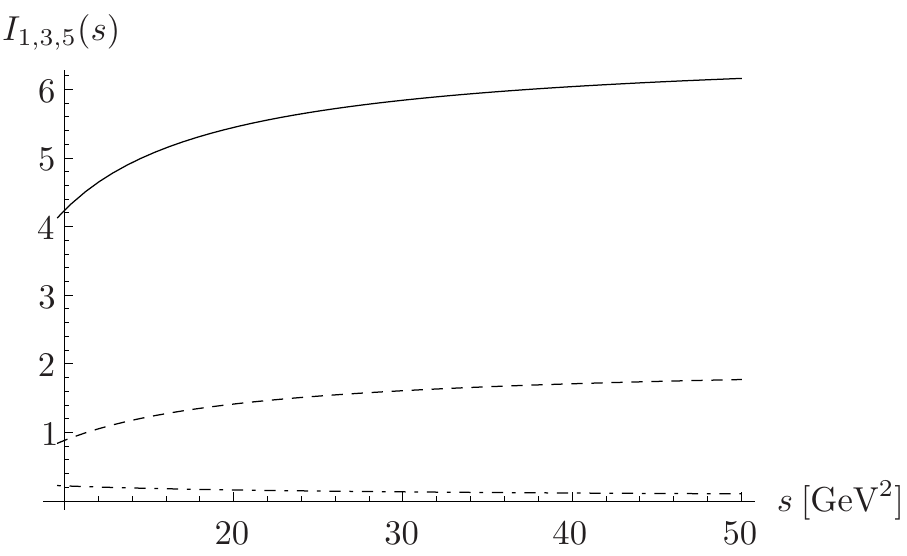}
\caption{Functions $I_1(s)$ (see eq.~\eqref{I1I3}) (solid line), $I_3(s)$ (dashed line), and the correction $I_5(s)\,l^{(\Lambda)}_p\delta(s)/(m_p \beta_\Lambda)$ (dash-dotted line).}
\label{fig:I135}
\end{figure}

First, we change variables to rationalize our integral:
\begin{equation}\label{changeofvar}
\delta\theta_1^{(0)}=\delta(s) \sqrt{1-R^2} \cos \widetilde{\Phi}, \quad
\delta \phi_1^{(0)}= \frac{\delta(s) \sqrt{1-R^2} \sin
\widetilde{\Phi}}{\sqrt{\sin \theta_1^{(0)} \sin\left(\theta_1^{(0)}-\delta(s)
\sqrt{1-R^2} \cos \widetilde{\Phi}\right)}}.
\end{equation}
The variables $R$ and $\widetilde{\Phi}$ are the polar coordinates for the
integration region which is now a unit disk
\begin{equation}\label{measure}
\int_{g \geq g_{th}} \sin \theta \dd\theta \dd\phi \Big(\ldots\Big) \approx \delta^2(s) \int^{2\pi}_{0} \dd \widetilde{\Phi} \int^1_0 \dd R R \sqrt{\frac{\sin\left(\theta_1^{(0)} - \delta(s) \sqrt{1-R^2} \cos \widetilde{\Phi} \right)}{\sin \theta_1^{(0)}}}\Big(\ldots\Big).
\end{equation}
Here the ellipsis stands for the integrand defined in eq.~\eqref{sigma1mt}, where we expand $g$ and $g'$:
\begin{equation}\label{approximations}
\begin{split}
g & = 1-\frac{1}{2}\left(\delta\theta_1^{(0)\,2}+ \sin\theta^{(0)}_1
\sin\left(\theta^{(0)}_1-\delta\theta_1^{(0)}\right)
\delta \phi_1^{(0)\,2}\right)+ {\cal O}\left(\delta^4(s)\right), \\
g' & = -\delta\theta_1^{(0)}+\frac{1}{4}
\sin\left(2\theta_1^{(0)}\right)\delta\phi_1^{(0)\,2} + 
{\cal O}\left(\delta^3(s)\right),
\end{split}
\end{equation}
and rewrite them through $R$ and $\widetilde{\Phi}$:
\begin{equation}\label{approximations1}
\begin{split}
g & = 1-\frac{\delta^2(s)}{2}(1-R^2)+{\cal O}\left(\delta^4(s)\right), \\
g' & = - \delta(s) \sqrt{1-R^2} \cos \widetilde{\Phi} +\frac{\delta^2(s)}{2} (1-R^2) \cot\theta_1^{(0)} \sin^2 \widetilde{\Phi}+{\cal O}\left(\delta^3(s)\right).
\end{split}
\end{equation}
These expansions allow us to obtain the following relations with 
${\cal O}\left(\delta^2(s)\right)$ accuracy:
\begin{equation}\label{approximations2}
\begin{split}
\frac{l_p^{(0)}}{l^{(\Lambda)}_p} & \approx\gamma_\Lambda 
\frac{\left(\epsilon_p^{(\Lambda)} \beta_\Lambda /l^{(\Lambda)}_p \pm R \right)}{1+\left(1-R^2\right) l^{(\Lambda)\,2}_p /m^2_p}, \\
\frac{l_p^{(0)}}{\sqrt{s}/2-\epsilon_p^{(0)}} & \approx
\frac{\left(\beta_\Lambda \epsilon_p^{(\Lambda)} \pm l^{(\Lambda)}_p R
\right)}{\epsilon_p^{(\Lambda)}\left(\epsilon_p^{(\Lambda)}m_\Lambda /m^2_p-1\right)\mp \beta_\Lambda l^{(\Lambda)}_p R - m_\Lambda l^{(\Lambda)\,2}_p R^2 /m^2_p}, \\ 
\gamma_\Lambda \frac{g\, l_p^{(0)}-
\beta_\Lambda \epsilon_p^{(0)}}{l^{(\Lambda)}_p} & \approx \frac{\pm \, R
- \left(1-R^2\right) \epsilon_p^{(\Lambda)}l^{(\Lambda)}_p/(m^2_p\beta_\Lambda)
}{1+\left(1-R^2\right) l^{(\Lambda)\,2}_p /m^2_p}. 
\end{split}
\end{equation}
Next, we integrate eq.~\eqref{sigma1mt} with respect to $\widetilde{\Phi}$ and $R$ and get the cross section in the leading and next-to-leading order in $\delta(s)$:
\begin{equation}\label{sigma1int}
\begin{split}
\frac{d\sigma}{d \Omega^{(0)}_1} & = 
\frac{\mathcal{B}(\Lambda \to p\pi^-)\,s\,m_\Lambda \beta_\Lambda \,\alpha^{\xi}_\jpsi\alpha_{g}}
{2\,l^{(\Lambda)}_p\Big[(s-m^2_\jpsi)^2+m^2_\jpsi \Gamma^2_\jpsi\Big]} 
\frac{\left|G^\psi_M\right|^2}{1+\alpha} \delta^2(s) 
\Biggl\{\left(1+\alpha \cos^2 \theta_1^{(0)}\right)I_1(s)+ \\ 
 & + \xi \alpha_1 \Biggl[(1+\alpha) I_3(s)+ \sqrt{1-\alpha^2} \cosdphi \left(\frac{l^{(\Lambda)}_p}{m_p \beta_\Lambda}\delta(s)\right) I_{5}(s) \Biggr]\cos \theta_1^{(0)} +{\cal O}\left(\delta^2(s)\right)\Biggr\},
\end{split}
\end{equation}
where $I_1(s)$, $I_3(s)$, $I_5(s)$ are dimensionless functions of $s$:
\begin{equation}\label{I1I3}
\begin{split}
I_1(s) & = \frac{1-\kappa}{2\beta^2_p \kappa^2}
\left(\frac{\beta^2_\Lambda - 2 \kappa} {\tau(\beta_\Lambda)}L_1(\beta_\Lambda)+\beta_\Lambda 
L_2(\beta_\Lambda)\right), \\
I_3(s) & = \frac{(1-\kappa)(\mu^2-\kappa)}{2 \beta^3_p \kappa^3
\beta_\Lambda}\left(\frac{\beta^2_\Lambda -2 \kappa
}{\tau(\beta_\Lambda)}L_1(\beta_\Lambda)+\beta_\Lambda 
L_2(\beta_\Lambda)\right)-\frac{2 \mu^2}{\beta^3_p \kappa
^2}\log\left(\frac{\kappa}{\mu}\right), \\ 
I_5(s) & = \frac{(1-\kappa)\mu^2}{2 \beta^5_p \kappa^5}
\biggl[\beta_\Lambda \frac{\beta^2_\Lambda(\kappa-1)- 4\kappa^2-\mu^2+\kappa(3+2\mu^2)}{\tau(\beta_\Lambda)}L_1(\beta_\Lambda)+ \\ 
& \quad +\left(\beta^2_\Lambda (\kappa-1)-\mu^2+\kappa\right)L_2(\beta_\Lambda)
\biggr] + \frac{2\mu^4}{\beta^5_p \kappa^4} \log\left(\frac{\kappa}{\mu}\right)-\frac{\mu^2}{\beta^3_p \kappa^3}.
\end{split}
\end{equation}
Here we introduced 
\begin{equation}
 \begin{split}
    \kappa&=\epsilon_p^{(\Lambda)} m_\Lambda /m^2_p \approx 1.196,\\
       \mu&=m_\Lambda /m_p \approx 1.189,\\
   \beta_p&=\beta_p^{(\Lambda)}=l^{(\Lambda)}_p/\epsilon_p^{(\Lambda)}= \sqrt{1-\mu^2/\kappa^2}\approx 0.107.
 \end{split}
\end{equation}
Moreover, $I_1$, $I_3$, and $I_5$ depend on $s$ through $\beta_\Lambda =\sqrt{1-4m_\Lambda^2/s}$ and the following functions of $\beta_\Lambda$:
\begin{equation}\label{tauL1L2}
 \begin{split}
  \tau(\beta_\Lambda)&=\sqrt{\beta^2_\Lambda +4\kappa(\kappa-1)}, \\ L_1(\beta_\Lambda)&=\log\left[\frac{(\kappa^2-\mu^2)(\beta_\Lambda -\tau)^2-4\kappa^2 (\kappa -1)^2}{(\kappa^2-\mu^2)(\beta_\Lambda +\tau)^2-4\kappa^2 (\kappa -1)^2}\right], \\
  L_2(\beta_\Lambda)&=\log\left[
  \frac{(\kappa-\mu^2)^2-\beta^2_\Lambda (\kappa^2-\mu^2)}{\kappa^2(\kappa-1)^2}\right].
 \end{split}
\end{equation}
For extremely large energies the functions $I_{1,3,5}$ saturate:
$I_1\to 6.638$, $I_3\to 2.026$, $I_5\to 28.801$,
but the correction $\left(l^{(\Lambda)}_p/(m_p \beta_\Lambda)\,\delta(s)\right) I_5(s)$ vanishes, as is shown in figure~\ref{fig:I135}.
%===============================================================
%EndExpansion
For the energy of the $\jpsi$-resonance, $s_\jpsi=m^2_\jpsi=\left(3096.9\ \rm{MeV}\right)^2$, one gets the following values:
\begin{equation}\label{values}
 I_1(s_\jpsi)\approx4.122, \qquad I_3(s_\jpsi)\approx0.839,
 \qquad \left(\frac{l^{(\Lambda)}_p}{m_p
 \beta_\Lambda}\delta(s_\jpsi)\right) I_5(s_\jpsi)\approx
 0.223
\end{equation}
with the accuracy of $\delta^2(s_\jpsi)\approx 0.01$.

From eq.~\eqref{sigma1int} we have for $s=s_\jpsi$
\begin{equation}\label{sigmap}
\begin{split}
\frac{d\sigma}{d\Omega^{(0)}_1} & = \mathcal{B}(\Lambda\to 
p\pi^-) \mathcal{B}(\jpsi \to e^+e^-) \mathcal{B}(\jpsi \to 
\Lambda \lambar) \frac{\alpha^{\xi}_\jpsi}{\alpha_\jpsi} 
\frac{9 m_\Lambda}{l^{(\Lambda)}_p m^2_\jpsi} 
\frac{I_1(s_\jpsi)}{(3+\alpha)}\times \\ 
&\quad \times \delta^2(s_\jpsi)\Biggl\{1+\alpha\cos^2\theta_1^{(0)} + \xi\alpha_1\cos\theta_1^{(0)}\Biggl[(1+\alpha) \frac{I_3(s_\jpsi)}{I_1(s_\jpsi)}+\\
& \qquad +\sqrt{1-\alpha^2}\cosdphi \left(\frac{l^{(\Lambda)}_p}
{m_p \beta_\Lambda}\delta(s_\jpsi)\right) \frac{I_{5}(s_\jpsi)}{I_1(s_\jpsi)} \Biggr] + {\cal O}\left(\delta^2(s_\jpsi)\right)\Biggr\},
\end{split}
\end{equation}
which gives the numerical result presented in eq.~\eqref{eq:num_cm}.
%  Using eqs.~\eqref{alphas}, \eqref{values} and the following parameters from ref.~\cite{pdg}:
% \begin{equation}\label{constants}
% \begin{split}
% & m_\jpsi = 3096.9\ \mathrm{MeV}, \quad
% m_\Lambda = 1115.7\ \mathrm{MeV}, \quad
% m_p = 938.2721\ \mathrm{MeV}, \quad
% m_\pi = 139.57\ \mathrm{MeV}, \\
% & \mathcal{B}(\Lambda \to p\pi^-) = 63.9 \%\,,\quad
% \mathcal{B}(\jpsi \to e^+e^-) = 5.971 \%\,,\quad
% \mathcal{B}(\jpsi \to \Lambda \lambar) = 1.89 \times 10^{-3},
% \end{split}
% \end{equation}
% we get the numerical result presented in eq.~\eqref{eq:num_cm}.Targeted SCT luminosity of~$10^{35}~\lumi$ will provide about~$10^{12}$~$\jpsi$ mesons detected during a~$10^7~\mathrm{s}$ long period of data taking. Presume the data set is divided into three equal parts containing~$N_0\approx 3\cdot 10^{11}$ events each, corresponding to 1) beam with $+|\pe|$ average polarization, 2) beam with $-|\pe|$ average polarization, and 3) unpolarized beam. Assuming~$\p

%%%%%%%%%%%%%%%%%%%%%%%%
%%%%  BIBLIOGRAPHY  %%%%
%%%%%%%%%%%%%%%%%%%%%%%%

\end{document}